**Fast explicit dynamics finite element algorithm for transient heat transfer**

Jinao Zhang, Sunita Chauhan

Department of Mechanical and Aerospace Engineering, Monash University, Wellington Rd, Clayton, Melbourne, VIC 3800, Australia

**Abstract.** This paper presents a novel methodology for fast simulation and analysis of transient heat transfer problems. The proposed methodology is suitable for real-time applications owing to (i) establishing the solution method from the viewpoint of computationally efficient explicit dynamics, (ii) proposing an element-level thermal load computation to eliminate the need for assembling global thermal stiffness matrix, leading to (iii) an explicit formulation of nodal temperature computation to eliminate the need for iterations anywhere in the algorithm, (iv) pre-computing the constant matrices and simulation parameters to facilitate online calculation, and (v) utilising computationally efficient finite elements to efficiently obtain thermal responses in the spatial domain, all of which lead to a significant reduction in computation time for fast run-time simulation. The proposed fast explicit dynamics finite element algorithm (FED-FEM) employs nonlinear thermal material properties, such as temperature-dependent thermal conductivity and specific heat capacity, and nonlinear thermal boundary conditions, such as heat convection and radiation, to account for the nonlinear characteristics of transient heat transfer problems. Simulations and comparison analyses demonstrate that not only can the proposed methodology handle isotropic, orthotropic, anisotropic and temperature-dependent thermal properties but also satisfy the standard patch tests and achieve good agreement with those of the commercial finite element analysis packages for numerical accuracy, for three-dimensional (3-D) heat conduction, convection, radiation, and thermal gradient concentration problems. Furthermore, the proposed FED-FEM algorithm is computationally efficient and only consumes a small computation time, capable of achieving real-time computational performance, leading to a novel methodology suitable for real-time simulation and analysis of transient heat transfer problems. The source code is available at https://github.com/jinaojakezhang/FEDFEM.

**Keywords:** Transient heat transfer; real-time computation; explicit time integration; finite element analysis; nonlinear heat transfer analysis.

## 1. Introduction

Transient heat transfer analysis is of great interest in many engineering applications [1-3]. The solution methods to the transient heat transfer problems are mainly based on the analytical and numerical methods. The analytical method can provide exact solutions but only exists for regular geometric shapes with isotropic and homogeneous material properties and simple boundary conditions [4], and hence its application in complex heat transfer problems are limited. The numerical methods are flexible and effective for handling complex transient heat transfer problems. The most commonly used numerical methods are the finite element method (FEM) [5, 6], finite difference method (FDM) [7, 8], finite volume method (FVM) [9, 10], boundary element method (BEM) [11, 12], and meshless methods [13, 14], among which the FEM is most widely utilised due to its flexibility in handling complex geometries, material properties, and boundary conditions [15]. In FEM, an approximate discrete representation of the original physical problem is obtained by subdividing the problem domain into a number of elementary building components called the finite elements, forming a finite element mesh that conforms to the problem domain [16]. The differential equation governing the physical behaviour of the problem is solved based on the variational or energy principles of virtual work, and it is approximated with respect to each element. The individual equations of the finite elements are assembled into a large system of equations, from which the field variable unknowns are obtained as solutions. Despite the physical accuracy of FEM, it is computationally expensive. The literature reported that simulating a transient heat transfer problem using standard FEM resulted in high computation times ranging from a few minutes to several hours [17]; for instance, it was reported that a simulation of an 8-minute thermal ablation treatment took around 6.0 hours [18].

Many engineering applications, such as the virtual-reality or augmented-reality-based applications, require realistic and real-time solution of transient heat transfer problems [19]; however, most of the developed numerical methods [20-24] are focused on the former (numerical accuracy, convergence, and stability), rather than the computation time to satisfy the conflicting requirements. Heat transfer problems are complex in terms of material characteristics which may be time/temperature-dependent and heat transfer behaviours such as heat convection and radiation, leading to nonlinear characteristics of transient heat transfer problems [22]. However, owing to the involvement of both nonlinear thermal properties and nonlinear thermal boundary conditions, the use of the traditional FEM-based solution method is computationally expensive. It requires a long computation time for a simulation of a transient nonlinear heat transfer problem, since the nonlinear system of equations needs to be iteratively solved at each time step [25], leading to the difficulty in achieving real-time computational performance. To facilitate the computational efficiency of FEM for transient heat transfer problems, Feng et al. [26] employed the combined approximations (CA) method, where global and local approximations are combined for an efficient solution; however, the formulation of CA is not based on the rigorous mathematical proof [2]. Later, Feng et al. [2] also studied a reduced-basis method obtained through a rigorous theoretical proof of model order reduction. The essential idea of the reduced order modelling is to employ a set of global basis, that is, in a statistical sense, the best suited to reproduce the complete problem, by which the full system response is projected onto a smaller dimensional subspace, leading to a reduction in number of degrees of freedom for fast numerical calculation [27, 28]. Also based on the concept of reduced order modelling, Ding et al. [25] presented an isogeometric independent coefficients (IGA-IC) reduced order method to improve the computational performance of FEM. Similarly, Wang et al. [29] investigated the CA and IC methods for efficiently obtaining the steady temperature fields without the need for solving the global system of equations. Despite the improvement on numerical efficiency made by the model reduction methods, the accuracy and computational effort of the algorithm are dependent on the number of reduced-basis chosen [2], and there is an inevitable loss in energy due to the projection from a full system space to a smaller subspace [30]. In general, the two conflicting requirements (realistic and real-time) raise technical challenges for simulation and analysis of transient heat transfer problems for real-time-based engineering applications. Performance improvement on one requirement is mainly achieved by performance sacrifice on the other.

This paper presents a new methodology for fast simulation and analysis of transient heat transfer problems, and it is suitable for real-time applications. For fast and efficient numerical updates in the run-time computation, the proposed methodology formulates the transient heat transfer problems as computationally efficient explicit dynamics in the temporal domain, and the thermal responses are obtained directly at nodes based on the finite element discretisation in the spatial domain. Without the need for iterative solutions for unknown field variables, the proposed methodology devises an efficient explicit formulation for nodal temperature calculation, where nodal thermal loads are determined at the element-level from the finite elements sharing the node. With this formulation, it allows for pre-computation of several simulation parameters, such as the time-stepping constants and nodal spatial derivatives. By further utilising computationally efficient finite elements, the proposed methodology can produce a significant reduction in computation time for fast run-time computation. The proposed FED-FEM can accommodate various thermal material properties and handle nonlinear characteristics involved in transient heat transfer problems. The computational benefits of the proposed solution procedure are examined, and an assessment of numerical errors is presented. It is demonstrated that real-time computations become possible for the proposed FED-FEM algorithm while maintaining numerical accuracy, even in the cases involving nonlinear thermal properties, nonlinear boundary conditions, and concentrated thermal gradients, leading to an efficient and accurate numerical algorithm for conducting transient heat transfer analyses for real-time applications.

The remaining of this paper is organised as follows: Section 2 first gives a brief overview of the transient heat transfer problem. Section 3 discusses the proposed methodology, including explicit time-stepping procedure, explicit nodal temperature formulation, element-wise nodal thermal load computation, and numerical stability as well as the algorithm. Section 4 evaluates the performance of the proposed FED-FEM in terms of numerical convergence, accuracy and computational efficiency, and discussions are presented in Section 5. Finally, this paper concludes in Section 6 with future improvements of the work.

## 2. Transient heat transfer problems

The transient heat transfer problems in a time-continuous spatial domain can be physically interpreted as a statement of heat flow equilibrium based on the law of conservation of energy [4]. The strong-form equation governing transient heat transfer problems in the system $\Omega$ can be written as

$$\rho c(T)\frac{\partial T(\mathbf{x},t)}{\partial t}=\frac{\partial}{\partial x_i}\left(k_{ij}(T)\frac{\partial T(\mathbf{x},t)}{\partial x_j}\right)+Q(\mathbf{x},t),\qquad t>t_0,\mathbf{x}\in\Omega \tag{1}$$

where $\rho$ is the material density; $c(T)$ the specific heat capacity and is, in general, a temperature-dependent variable in the function of temperature $T$ which is the field variable at spatial point $\mathbf{x}\in\Omega$ and varies with time $t$; $x_i$ the $i$th component of the spatial coordinates at point $\mathbf{x}=(x_1,x_2,\dots,x_n)$ where $n$ denotes the dimension of the problem; $k_{ij}(T)$ the thermal conductivity in the direction of the $i$th component of the spatial coordinates and $j$th component of the temperature gradient, and it is, in general, also a temperature-dependent variable in the function of temperature $T$; for orthotropic conductivity $k_{ij}(T)=0$ for $i\neq j$, i.e. $\frac{\partial}{\partial x_i}\left(k_{ii}(T)\frac{\partial T(\mathbf{x},t)}{\partial x_i}\right)$, and for isotropic conductivity $k_{ii}(T) = k(T)$, i.e. $k(T)\frac{\partial^2 T(\mathbf{x},t)}{\partial {x_i}^2}$; $Q(\mathbf{x}, t)$ the net heat flows at the spatial point $\mathbf{x}$ and varies with time $t$; and $t_0$ denotes the initial time.

The field problem of the transient heat transfer can be solved by the Galerkin weak-form finite element method via the matrix form of the discretised transient heat transfer problem and the associated initial and boundary conditions. The problem domain is divided into a finite number of subdomains or elements, forming a finite element mesh in one, two, and three dimensions that conforms to the problem domain, to approximate the governing equations of the transient heat transfer. The field variables are interpolated by separate functions that have common values at nodes of the element system [4]. Individual equations of the finite elements are assembled into a large system of equations, from which the unknown field variable values (nodal temperatures) are obtained as solutions. In the finite element formulation, problem domain of arbitrary shapes can be accommodated without difficulty, and mesh sizes can be varied by considering the numerical accuracy required and computational resources available. Material properties can be integrated into the parametric constitutive equations and can vary over elements. Boundary conditions can be handled directly and accounted for in the global system of equations for elements subjected to boundary conditions. The governing equation for transient heat transfer problems using the standard Galerkin weak-form finite element formulation is expressed in the following matrix form

$$\mathbf{C}(T)\frac{\partial \mathbf{T}(\mathbf{x},t)}{\partial t} = \mathbf{K}(T)\mathbf{T}(\mathbf{x},t) + \mathbf{Q}(\mathbf{x},t), \qquad t > t_0, \mathbf{x} \in \Omega \tag{2}$$

where $\mathbf{C}(T)$ and $\mathbf{K}(T)$ are the thermal mass (mass and specific heat) and thermal stiffness (conductivity) matrices, respectively; $\mathbf{T}(\mathbf{x}, t)$ the vector of unknown nodal temperatures; and $\mathbf{Q}(\mathbf{x}, t)$ the vector of net nodal heat flows; the initial condition at $t_0$ is $T(\mathbf{x}, 0) = T_0$ which denotes the initial temperature of the system.

The boundary $\Gamma$ in the heat transfer problem can be decomposed into the following five different boundary types: $\Gamma = \Gamma_E + \Gamma_C + \Gamma_H + \Gamma_A + \Gamma_R$, and the boundary conditions are expressed as follows:

(i) Essential boundary condition

$$T(\mathbf{x},t) = T_\Gamma, \qquad t > t_0, \mathbf{x} \in \Gamma_E \tag{3}$$

where $T_\Gamma$ is the prescribed temperature on the Dirichlet boundary.

(ii) Convection boundary condition

$$-k_{ij}(T)\frac{\partial T(\mathbf{x},t)}{\partial x_j} n_i = -h(T(\mathbf{x},t) - T_a), \qquad t > t_0, \mathbf{x} \in \Gamma_C \tag{4}$$

where $h$ is the convection heat transfer coefficient; $n$ the outward normal direction on the surface boundary; and $T_a$ the ambient temperature on the Robin boundary.

(iii) Heat flux boundary condition

$$-k_{ij}(T)\frac{\partial T(\mathbf{x},t)}{\partial x_j} n_i = q_i(\mathbf{x},t), \qquad t > t_0, \mathbf{x} \in \Gamma_H \tag{5}$$

where $q_i(\mathbf{x}, t)$ is the prescribed heat flux on the Neumann boundary.

(iv) Adiabatic boundary condition

$$k_{ij}(T)\frac{\partial T(\mathbf{x},t)}{\partial x_j} n_i = 0, \qquad t > t_0, \mathbf{x} \in \Gamma_A \tag{6}$$

(v) Radiation boundary condition

$$-k_{ij}(T)\frac{\partial T(\mathbf{x},t)}{\partial x_j} n_i = \sigma\varepsilon[(T - T_z)^4 - (T_a - T_z)^4], \qquad t > t_0, \mathbf{x} \in \Gamma_R \tag{7}$$

where $\sigma$ is the Stefan-Boltzmann constant ($5.67 \times 10^{-8}\ Wm^{-2}K^{-4}$) for radiation; $\varepsilon$ the emissivity of the radiant surface; $T_z$ the value of absolute zero temperature; and $T_a$ the ambient temperature.

## 3. Fast explicit dynamics finite element algorithm

As mentioned previously, fast and efficient solution methods to the transient heat transfer problems are of great importance in many practical engineering problems. To this end, the proposed fast explicit dynamics finite element algorithm addresses fast computation of transient heat transfer problems via the (i) explicit dynamics in temporal domain, (ii) element-level thermal load computation, (iii) explicit formulation for unknown nodal temperatures, (iv) pre-computation of simulation parameters, and (v) computationally efficient finite elements. Several simulation parameters, such as the system constants for time-stepping and temperature gradient matrix, can be pre-computed, leading to an efficient time-stepping procedure. It allows the straightforward treatment of nonlinearities and can accommodate non/linear thermal properties and boundary conditions. Different from standard FEM, a time step is performed directly without the need for numerical iterations for solutions of the nonlinear system of equations. The computational cost at each time step is low, and the independent equations for individual nodal unknown temperatures can be computed independently, permitting parallel implementation of the proposed algorithm for fast solutions of transient heat transfer problems.

### 3.1 Explicit time integration of the discretised equation

The dynamics of transient heat transfer are commonly obtained via numerical time integration schemes such as the explicit [15, 31] and implicit [3, 22] integrations; the time-dependent variables are discretised using a finite difference technique via a time increment to estimate the continuous field variables in the next time point. By choosing different finite difference techniques, such as the forward or backward finite difference estimates, an explicit or implicit integration scheme can be obtained [32]. The implicit scheme uses the backward finite difference estimation; variables in the future state are determined by the variables both at the current and future states, leading to a system of equations where unknown state variable values are implicitly given as solutions. The implicit integration is unconditionally stable for any arbitrarily chosen time step [33], with limitations only due to considerations of numerical convergence and computational accuracy. Despite its unconditional stability, the implicit integration is computationally more expensive than the explicit counterpart. It requires a solution of a nonlinear system of equations at each time step, which is usually solved by an iterative method such as the Newton-Raphson method through a sequence of solutions of linear equations. As reported, the computation time of one iteration step by the implicit integration is at least one order of magnitude larger than that by the explicit integration [34].

To achieve fast solutions of the discretised transient heat transfer equation, it requires an efficient numerical scheme for integrations in the temporal domain. The explicit integration is easy to implement and computationally efficient, since variables in the future state are obtained explicitly based on the current state of known values only, without the need for inversion of the stiffness matrix at each time step [15]. It is also well suited for distributed parallel computing since mass lumping techniques may be employed, by which the global system of equations can be split into independent equations for individual nodes, allowing each nodal equation to be assigned to a processor in the parallel computer to perform calculations independently.

Using the first-order explicit forward time integration scheme, the temporal derivative of the continuous field temperature can be written as

$$\frac{\partial T(\mathbf{x},t)}{\partial t} = \frac{T(\mathbf{x},t+\Delta t) - T(\mathbf{x},t)}{\Delta t} \tag{8}$$

where $\Delta t$ is the time step.

By substituting Eq. (8) into the spatially discretised matrix form of transient heat transfer (Eq. (2)), yielding

$$\mathbf{C}(T)\left(\frac{\mathbf{T}(\mathbf{x},t+\Delta t) - \mathbf{T}(\mathbf{x},t)}{\Delta t}\right) = \mathbf{K}(T)\mathbf{T}(\mathbf{x},t) + \mathbf{Q}(\mathbf{x},t) \tag{9}$$

and it can be further arranged into

$$\left(\frac{\mathbf{C}(T)}{\Delta t}\right)\mathbf{T}(\mathbf{x},t+\Delta t) = \sum_{e}\mathbf{F}_e(\mathbf{x},T,t) + \mathbf{Q}(\mathbf{x},t) + \left(\frac{\mathbf{C}(T)}{\Delta t}\right)\mathbf{T}(\mathbf{x},t) \tag{10}$$

where

$$\sum_{e}\mathbf{F}_e(\mathbf{x},T,t) = \mathbf{K}(T)\mathbf{T}(\mathbf{x},t) = \mathbf{F}(\mathbf{x},T,t) \tag{11}$$

where $\mathbf{F}_e(\mathbf{x},T,t)$ are the components due to conduction in element $e$ of the global nodal thermal loads $\mathbf{F}(\mathbf{x},T,t)$. For a given element $e$, $\mathbf{F}_e(\mathbf{x},T,t)$ may be computed by

$$\mathbf{F}_e(\mathbf{x}, T, t) = \int_{V_e} \mathbf{B}(\mathbf{x})^T \mathbf{D}(T) \mathbf{B}(\mathbf{x})\, dV\, \mathbf{T}(\mathbf{x}, t) \tag{12}$$

where $V_e$ is the $e$th element volume, $\mathbf{B}(\mathbf{x})$ the temperature gradient matrix, and $\mathbf{D}(T)$ the thermal conductivity matrix.

By employing the lumped (diagonal) mass approximation while ensuring mass conservation, it leads to a diagonal thermal mass matrix; by subsequently multiplying field temperatures, it renders Eq. (10) an explicit formulation for the unknown field temperatures $\mathbf{T}(\mathbf{x}, t + \Delta t)$. Eqs. (10), (11), and (12) further imply that computations are performed at the element level, eliminating the need for assembling the thermal stiffness matrix $\mathbf{K}(T)$ and multiplying the temperature $\mathbf{T}(\mathbf{x}, t)$ for the entire model. Therefore, the computation cost at each time step is significantly lower in the proposed methodology compared to standard FEM for there is no need for forming the global system of equations and applying iterations for solution.

Assuming that temperatures $\mathbf{T}(\mathbf{x}, t)$ at the current time step are known, and that the nodal thermal loads $\mathbf{F}(\mathbf{x}, T, t)$ have been computed using Eq.(11), the explicit forward time integration renders an equation for obtaining nodal temperatures at the next time point, yielding

$$\mathbf{T}(\mathbf{x}, t + \Delta t) = \left(\frac{\Delta t}{\boldsymbol{\rho}\mathbf{c}(T)}\right)\left(\mathbf{F}(\mathbf{x}, T, t) + \mathbf{Q}(\mathbf{x}, t)\right) + \mathbf{T}(\mathbf{x}, t) \tag{13}$$

where $\boldsymbol{\rho}$ and $\mathbf{c}(T)$ are the lumped mass and specific heat capacity matrices.

By defining the diagonal coefficient matrix

$$\mathbf{A} = \frac{\Delta t}{\boldsymbol{\rho}} \tag{14}$$

Eq. (13) may be written as

$$\mathbf{T}(\mathbf{x}, t + \Delta t) = \left(\frac{\mathbf{A}}{\mathbf{c}(T)}\right)\left(\mathbf{F}(\mathbf{x}, T, t) + \mathbf{Q}(\mathbf{x}, t)\right) + \mathbf{T}(\mathbf{x}, t) \tag{15}$$

For each time step, the nodal temperatures are obtained by

$$T^{(k)}(\mathbf{x}, t + \Delta t) = \left(\frac{A^{(k)}}{c^{(k)}(T)}\right)\left(F^{(k)}(\mathbf{x}, T, t) + Q^{(k)}(\mathbf{x}, t)\right) + T^{(k)}(\mathbf{x}, t) \tag{16}$$

where $A^{(k)}$ and $c^{(k)}(T)$ are the diagonal entries in the $k$th row of the diagonal coefficient matrix **A** and diagonal specific heat capacity matrix $\mathbf{c}(T)$, respectively.

The nonlinear thermal properties are accounted for in the calculation of thermal heat capacity matrix $\mathbf{c}(T)$ and the thermal conductivity matrix $\mathbf{D}(T)$, and nonlinear boundary conditions are accounted for in the calculation of the net nodal heat flows $\mathbf{Q}(\mathbf{x}, t)$. It is worth noting that there is no need for iterations anywhere in the algorithm, and the coefficient matrix **A** may be pre-computed; hence, Eqs. (15) and (16) provides an efficient means for advancing temperatures in time. Furthermore, Eq. (16) states an explicit formulation of the unknown field temperatures $\mathbf{T}(\mathbf{x}, t + \Delta t)$; therefore, the global system of equations can be split into independent equations for individual nodes, permitting parallel implementation of the proposed algorithm to perform calculations independently.

### 3.2 Computation of element nodal thermal loads

The integral in Eq. (12) may be computed using Gaussian quadrature for nodal thermal loads. For fast and efficient computation of nodal thermal loads, the proposed FED-FEM employs the computationally efficient eight-node reduced integration hexahedral element and the four-node linear tetrahedral element for 3-D transient heat transfer problems.

For the eight-node reduced integration hexahedral element, it leads to the formulation for element nodal thermal loads

$$\mathbf{F}_e(\mathbf{x}, T, t) = 8\det(J)\mathbf{B}(\mathbf{x})^T \mathbf{D}(T) \mathbf{B}(\mathbf{x}) \mathbf{T}(\mathbf{x}, t) \tag{17}$$

where 8 is the constant integer for volume integral of the eight-node reduced integration hexahedral element (see [16] for details), and $J$ the pre-computed element Jacobian matrix defining the mapping between derivatives in global and element natural coordinates.

For the four-node linear tetrahedral element, it leads to the formulation for element nodal thermal loads

$$\mathbf{F}_e(\mathbf{x}, T, t) = V\mathbf{B}(\mathbf{x})^T\mathbf{D}(T)\mathbf{B}(\mathbf{x})\mathbf{T}(\mathbf{x}, t) \tag{18}$$

where $V$ is the element volume.

By defining the matrix $\mathbf{G}(\mathbf{x})$

$$\mathbf{G}(\mathbf{x}) = \begin{cases} 8\det(J)\mathbf{B}(\mathbf{x})^T & \text{reduced integration hexahedrons} \\ V\mathbf{B}(\mathbf{x})^T & \text{linear tetrahedrons} \end{cases} \tag{19}$$

The element nodal thermal loads can be computed by

$$\mathbf{F}_e(\mathbf{x}, T, t) = \mathbf{G}(\mathbf{x})\mathbf{D}(T)\mathbf{B}(\mathbf{x})\mathbf{T}(\mathbf{x}, t) \tag{20}$$

Since the temperature gradient matrix $\mathbf{B}(\mathbf{x})$ and the volume of an element may be pre-computed, the matrix $\mathbf{G}(\mathbf{x})$ may also be pre-computed; hence, at each time step the nodal thermal loads are updated by considering only the nonlinear thermal conductivity and variation of temperatures.

### 3.3 Formulation for linear transient heat transfer

Eqs. (15) and (20) account for the nonlinear thermal properties (temperature-dependent specific heat capacity and thermal conductivity) in the calculation of thermal heat capacity matrix $\mathbf{c}(T)$ and thermal conductivity matrix $\mathbf{D}(T)$. However, for linear transient heat transfer problems where temperature-independent specific heat capacity $\mathbf{c}$ and thermal conductivity $\mathbf{D}$ are employed, the diagonal coefficient matrix $\mathbf{A}$ in Eq. (14) and matrix $\mathbf{G}(\mathbf{x})$ in Eq. (19) may be further formulated as

$$\mathbf{A} = \frac{\Delta t}{\boldsymbol{\rho}\mathbf{c}} = \frac{\Delta t}{\mathbf{C}} \tag{21}$$

$$\mathbf{G}(\mathbf{x}) = \begin{cases} 8\det(J)\mathbf{B}(\mathbf{x})^T\mathbf{D}\mathbf{B}(\mathbf{x}) & \text{reduced integration hexahedrons} \\ V\mathbf{B}(\mathbf{x})^T\mathbf{D}\mathbf{B}(\mathbf{x}) & \text{linear tetrahedrons} \end{cases} \tag{22}$$

where $\mathbf{C}$ is the diagonal temperature-independent thermal mass matrix; similar to the nonlinear case, the coefficient matrix $\mathbf{A}$ and matrix $\mathbf{G}(\mathbf{x})$ may be pre-computed.

Subsequently, the temperature field at the next time point can be obtained by

$$\mathbf{T}(\mathbf{x}, t + \Delta t) = \mathbf{A}\big(\mathbf{F}(\mathbf{x}, t) + \mathbf{Q}(\mathbf{x}, t)\big) + \mathbf{T}(\mathbf{x}, t) \tag{23}$$

and the element nodal thermal loads can be computed by

$$\mathbf{F}_e(\mathbf{x}, t) = \mathbf{G}(\mathbf{x})\mathbf{T}(\mathbf{x}, t) \tag{24}$$

### 3.4 Stability analysis

Despite the computational efficiency and simple implementation, the explicit time integration is conditionally stable, meaning precautions need to be taken for chosen time steps to achieve stable simulations otherwise will explode numerically. The mathematical evaluation of the stability of an integration scheme can be conducted using the Dahlquist's test equation [35]

$$\dot{y} = \lambda y(t), \qquad y(t_0) = y_0 \tag{25}$$

with the analytic solution given by $y(t) = y_0 e^{\lambda t}$, where $\lambda$ is a constant.

An integration scheme that yields a bounded solution to Eq. (25) is said to be stable, and it is only bounded when $\Re\mathrm{e}(\lambda) \leq 0$. Using the explicit integration, Eq. (25) may be approximated as

$$\dot{y} = \frac{y(t + \Delta t) - y(t)}{\Delta t} = \lambda y(t) \tag{26}$$

Equation. (26) can be further arranged into

$$y(t+\Delta t) = \lambda \Delta t y(t) + y(t) = (1+\lambda\Delta t)^{(t+\Delta t)} y_0 \quad (27)$$

where the condition for $y(t+\Delta t)$ not to increase indefinitely is

$$|1+\lambda\Delta t| \leq 1 \quad (28)$$

It can be seen from Eq. (28) that the explicit integration is only conditionally stable, and the critical time step $\Delta t$ is determined by

$$\Delta t \leq \frac{2}{|\lambda|} \quad (29)$$

In finite element formulation of dynamic transient heat transfer problems, this means that the time step must meet the Courant-Friedrichs-Lewy (CFL) condition for numerical stability [36]. Mathematically, the maximum time step allowed for stable simulation is associated with the largest eigenvalue $\lambda_{max}$ of the thermal stiffness matrix and the thermal mass and damping values [15, 16, 31]. However, the computation of eigenvalues of heat transfer problem is computationally very expensive, especially when the thermal matrices are large. Some methods for prediction of the maximum eigenvalues and critical time steps have been reported in [31].

### 3.5 Description of FED-FEM algorithm

The algorithm of the proposed FED-FEM is illustrated in Fig. 1. The proposed FED-FEM algorithm consists of three main stages, which are the pre-computation stage, initialisation stage, and time-stepping stage. The numerical computations in each stage are given as follows:

(i) Pre-computation stage

1) Load object mesh and boundary conditions.
2) For each element: compute the determinant of Jacobian $\det(J)$ and temperature gradient matrix $\mathbf{B}(\mathbf{x})$; compute matrix $\mathbf{G}(\mathbf{x})$ based on Eqs. (19) or (22) considering the temperature-dependent/independent thermal conductivity.
3) Compute the diagonal thermal mass matrix; compute diagonal coefficient matrix **A** based on Eqs. (14) or (21) considering the temperature-dependent/independent specific heat capacity.

(ii) Initialisation stage

1) Initialise nodal temperature $T_0$ at $t=0$, and apply boundary conditions for the first time step $\Delta t$, such as the prescribed heat fluxes or/and temperatures:

(iii) Time-stepping stage

1) Loop over elements: compute element nodal thermal loads $\mathbf{F}_e(\mathbf{x}, T, t)$ based on Eq. (20) or $\mathbf{F}_e(\mathbf{x}, t)$ based on Eq. (24) considering temperature-dependent/independent thermal conductivity.
2) Perform a time step:
   2.1) Obtain net nodal thermal loads $\mathbf{F}(\mathbf{x}, T, t)$ or $\mathbf{F}(\mathbf{x}, t)$ at time $t$.
   2.2) Explicitly compute nodal temperatures using the forward explicit formulation (Eqs. (15) or (23)); the explicit computation for Eq. (15) is given in Eq. (16).
   2.3) Apply boundary conditions for the next time step $t+\Delta t$.

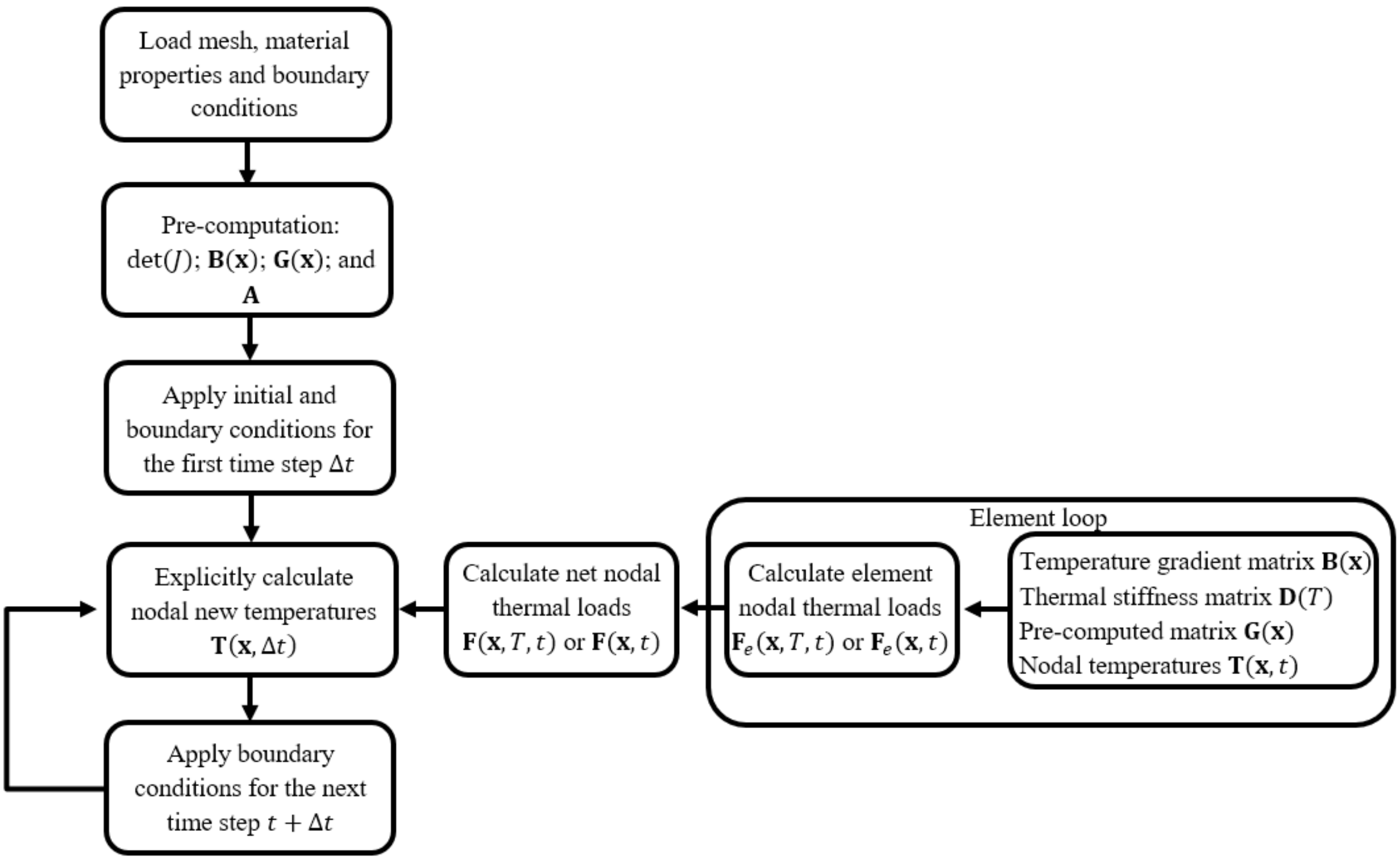

**Figure 1.** Proposed FED-FEM algorithm.

## 4. Performance evaluation

Simulations and comparison analyses are conducted to investigate the performance of the proposed FED-FEM algorithm. It first verifies the convergence of the proposed methodology by conducting standard patch tests, followed by verifications of numerical accuracy using 3-D heat conduction, convection, and radiation problems subject to different settings of thermal conductivity, specific heat, different types of boundary conditions, and different spatial meshes. It then investigates the computational performance by comparing computation times to those of the commercial finite element codes in terms of the isotropic, orthotropic, anisotropic and temperature-dependent thermal properties.

### 4.1 Patch tests

To verify a numerical method that can satisfy the requirement for convergence, it needs to pass the standard patch test [37]. As illustrated in Fig. 2, the patch tests are conducted in 3-D using a cube ($1\ m$ x $1\ m$ x $1\ m$) discretised into hexahedral and tetrahedral meshes for the eight-node reduced integration hexahedral element and the four-node linear tetrahedral element. A linear function of temperature $T = 200x + 100y + 200z$ is prescribed on the boundary nodes. To satisfy the patch test, temperatures at all interior nodes should follow exactly the same linear function as the imposed temperatures on the boundary nodes. The nodal temperatures (using three significant decimal digits of precision) are computed for all integration points at steady-state conditions and compared to the analytical solutions. Furthermore, the proposed FED-FEM is also verified against the commercial finite element analysis package, ABAQUS/CAE 2018 (license 6.20). Numerical errors in temperature for the hexahedral and tetrahedral meshes compared to the analytical solutions are 1.7e-03 and 18.3e-03, respectively, and they are 3.4158e-06 and 3.0278e-06, respectively, when compared to the ABAQUS solutions. The larger numerical error in the tetrahedral mesh may attributed to using the linear tetrahedral elements for large mesh distortion; however, both numerical errors are still on the order of e-03 and they are on the order of e-06 when compared to the ABAQUS solutions. Therefore, the proposed FED-FEM passes the patch tests with sufficient accuracy.

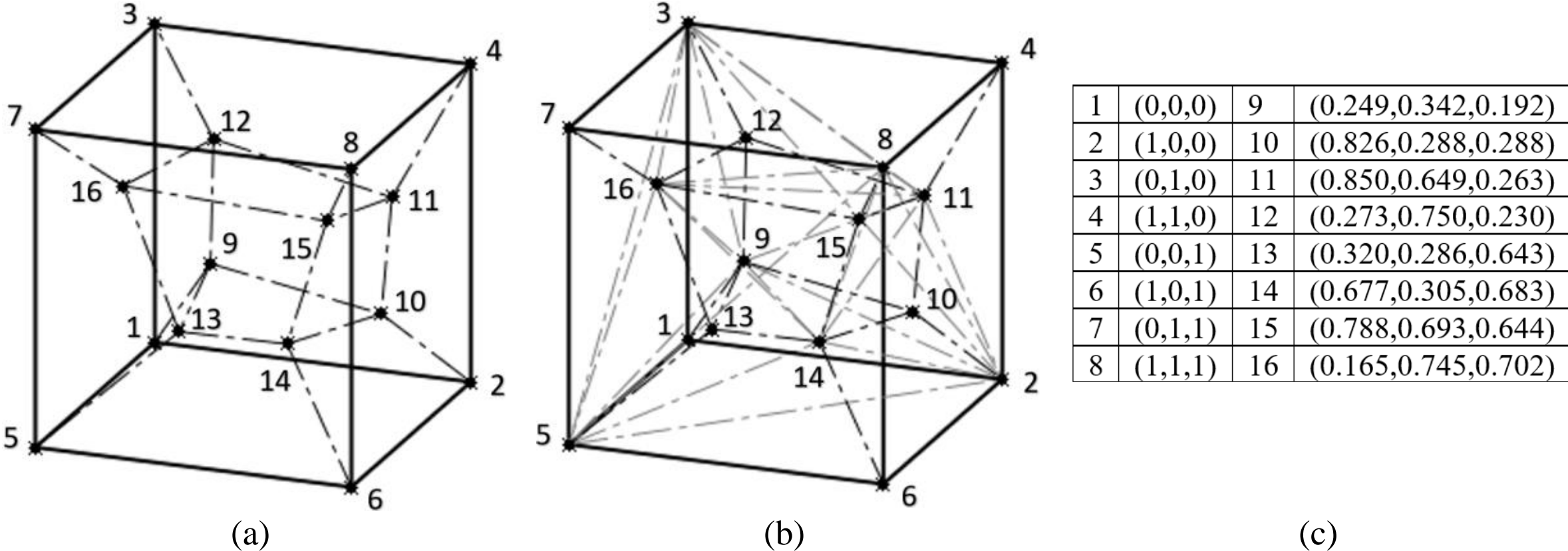

| | | | |
|---|---|---|---|
| 1 | (0,0,0) | 9 | (0.249,0.342,0.192) |
| 2 | (1,0,0) | 10 | (0.826,0.288,0.288) |
| 3 | (0,1,0) | 11 | (0.850,0.649,0.263) |
| 4 | (1,1,0) | 12 | (0.273,0.750,0.230) |
| 5 | (0,0,1) | 13 | (0.320,0.286,0.643) |
| 6 | (1,0,1) | 14 | (0.677,0.305,0.683) |
| 7 | (0,1,1) | 15 | (0.788,0.693,0.644) |
| 8 | (1,1,1) | 16 | (0.165,0.745,0.702) |

**Figure 2.** Patch tests of the proposed FED-FEM are performed on a (a) hexahedral mesh and (b) tetrahedral mesh with nodal coordinates given in (c).

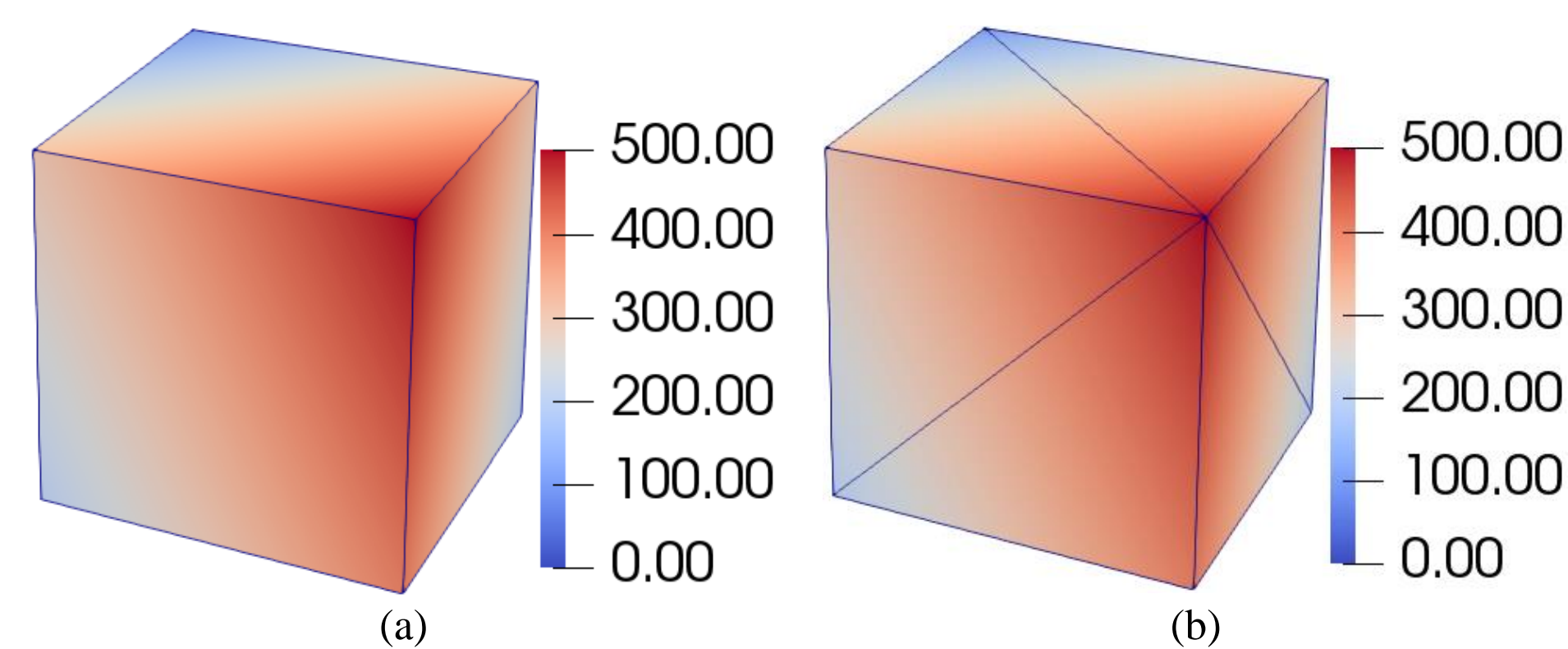

**Figure 3.** Temperature distribution of the patch tests on the (a) hexahedral mesh and (b) tetrahedral mesh.

## 4.2 Heat conduction

4.2.1 Isotropic, orthotropic, and anisotropic heat conduction over a 3-D cube

The proposed FED-FEM is verified against ABAQUS solutions by considering 3-D isotropic, orthotropic and anisotropic heat conduction problems. As illustrated in Fig. 4, the tested object is a cube ($0.1\ m$ x $0.1\ m$ x $0.1\ m$), and it is discretised into a tetrahedral mesh of four-node linear heat transfer tetrahedrons, leading to 40021 elements with 7872 nodes. Material properties are the constant isotropic thermal conductivity $k = 200\ W/(m \cdot ℃)$, medium density $\rho = 1000\ kg/m^3$, and specific heat $c = 2000\ J/(kg \cdot ℃)$. The initial temperature is $T_0 = 37$ ℃. The back face of the object is prescribed with a constant temperature $T_\Gamma = 37$ ℃ throughout the simulation, whereas a constant concentrated heat flux $q = 0.2\ W$ is applied to all nodes at the front, top, and two side faces; the adiabatic boundary condition is applied on the bottom face. The time step is $\Delta t = 0.01\ s$, and it can achieve a stable simulation.

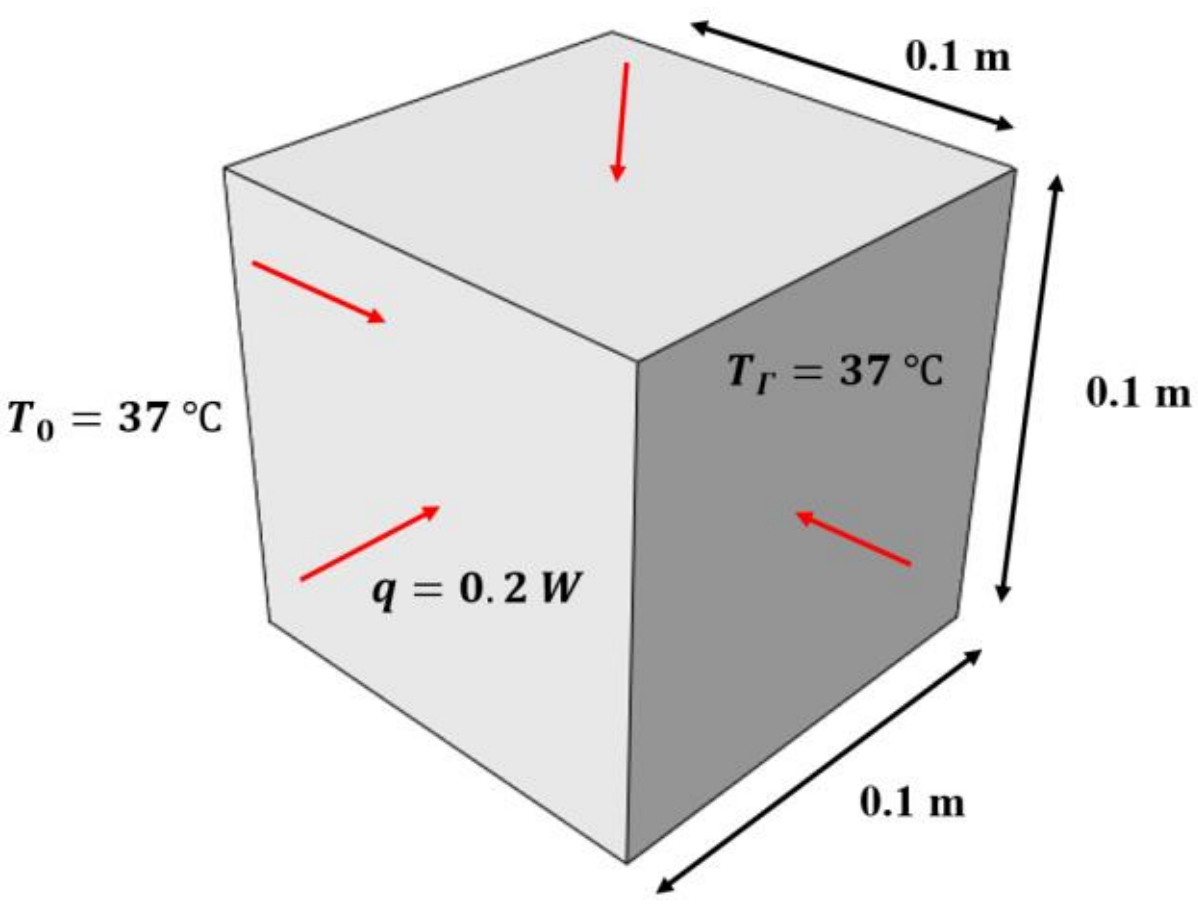

**Figure 4.** Geometry, initial, loading and boundary conditions of the tested cube: the geometry of the tested object is a cube ($0.1\ m$ x $0.1\ m$ x $0.1\ m$); the initial temperature is $T_0 = 37$ °C at all nodes; the back face is prescribed with a constant temperature $T_r = 37$ °C , the front, top, and two side faces are prescribed with a constant concentrated heat flux $q = 0.2\ W$ at all nodes on the face; and the adiabatic boundary condition is applied on the bottom face.

The steady state of the tested cube is computed using the proposed FED-FEM and compared to that of the ABAQUS ($\Delta T \leq 0.001$ °C). It can be seen from Fig. 5 that there is good agreement with those from the proposed FED-FEM compared with the ABAQUS solutions with the same set of cube mesh and simulation parameters. The statistical results of the comparison are presented in Table. 1.

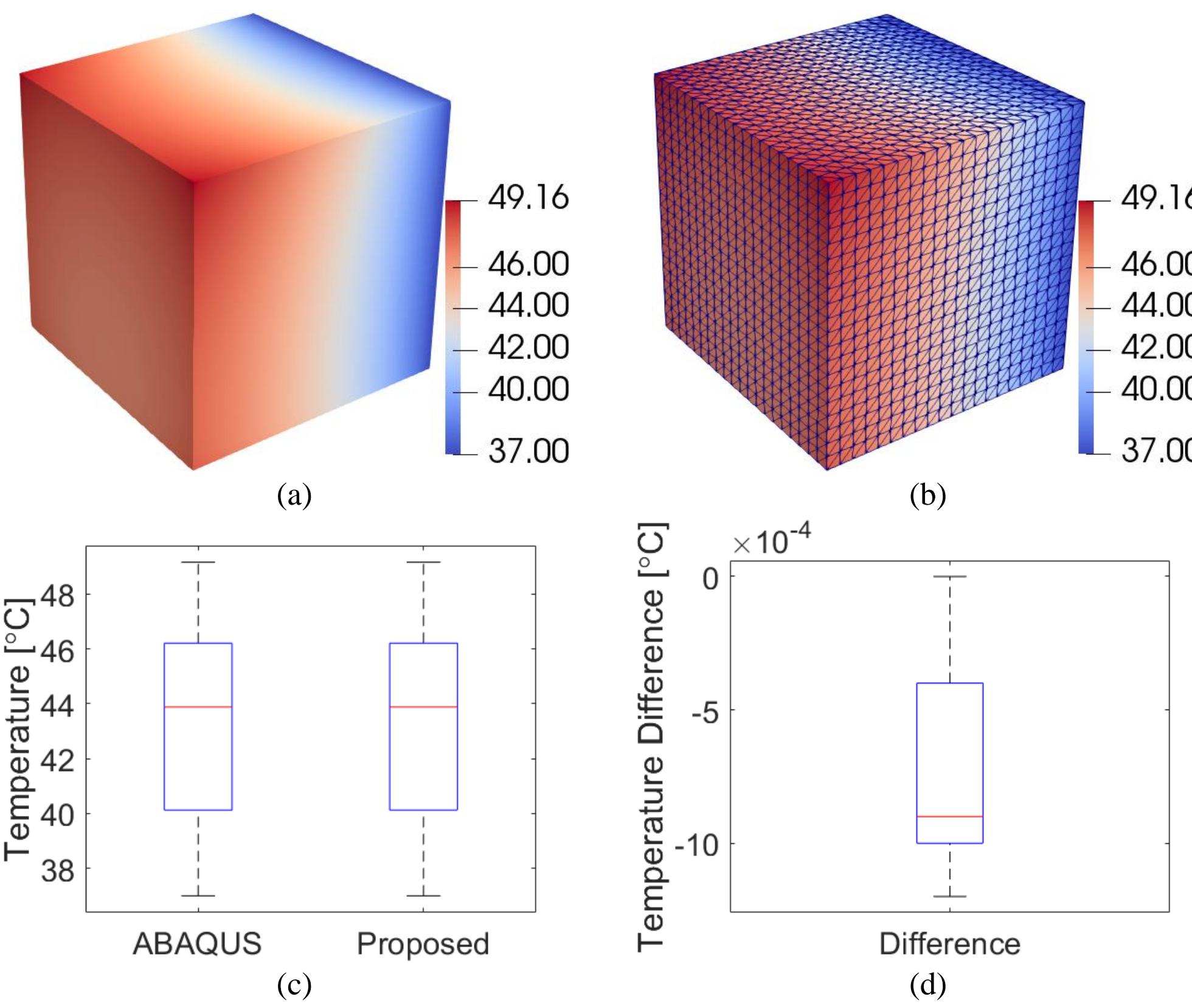

**Figure 5.** Comparison of the steady state of the tested cube: (a) temperature distribution computed by the proposed FED-FEM; (b) same temperature distribution with tetrahedral mesh displayed; (c) comparison of temperature at all nodes between ABAQUS and proposed FED-FEM; and (d) temperature differences for all nodes ($T^{ABAQUS} - T^{Proposed}$).

**Table. 1** Statistical results of the comparison in Fig. 5.

|  | Min (℃) | Max (℃) | Median (℃) | Q1 (℃) | Q3 (℃) |
|---|---|---|---|---|---|
| ABAQUS | 37 | 49.1563 | 43.8810 | 40.1272 | 46.2069 |
| FED-FEM | 37 | 49.1575 | 43.8819 | 40.1276 | 46.2080 |
| Difference | -0.0012 | 0 | -0.0009 | -0.0010 | -0.0004 |

Comparisons are also conducted for the transient heat conductions. The steady state is achieved at $t = 225.53\ s$ computed using ABAQUS, and the transient heat conduction comparisons are conducted at three intermediate time points $t = 56\ s$, $113\ s$ and $169\ s$. Fig. 6 illustrates comparisons of temperature between ABAQUS and proposed FED-FEM at the above time points and the associated temperature differences.

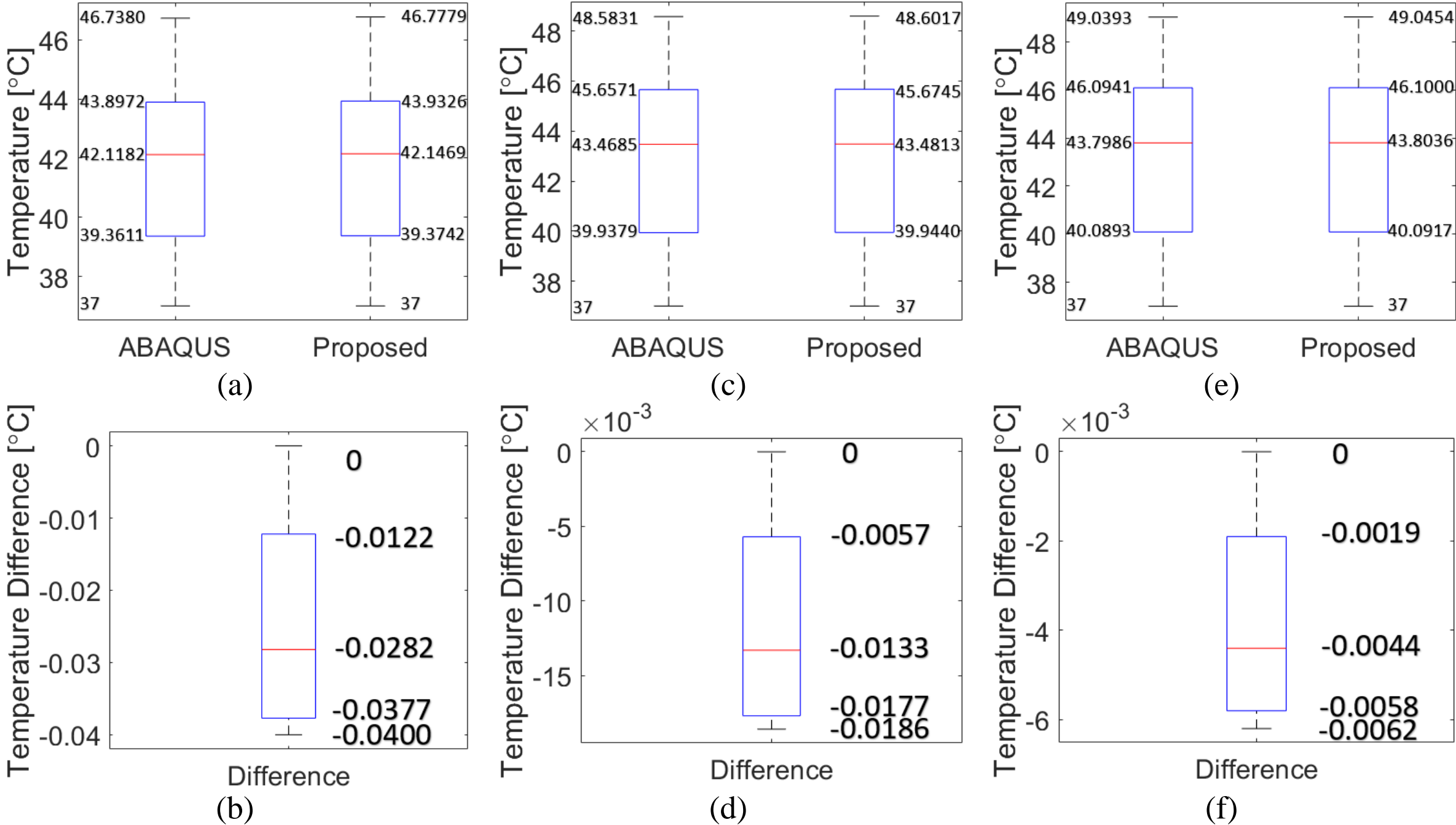

**Figure 6.** Comparison of transient heat conduction on the path to the steady state using the tested cube: comparisons of temperature at all nodes and temperature differences ($T^{ABAQUS} - T^{Proposed}$) at (a-b) $t = 56\ s$; (c-d) $t = 113\ s$; and (e-f) $t = 169\ s$.

Furthermore, the numerical error of the proposed FED-FEM is also investigated. The error used in this work is calculated using the following equation:

$$Error = \sqrt{\frac{\sum_{i=1}^{n}\left(T_i^{Abaqus} - T_i^{Proposed}\right)^2}{\sum_{i=1}^{n}\left(T_i^{Abaqus}\right)^2}} \tag{30}$$

The numerical errors of the proposed FED-FEM at the transient and steady states of the isotropic heat conduction are presented in Fig. 7. There is only marginal difference between the results obtained from the proposed FED-FEM and ABAQUS. Furthermore, It can also be seen that numerical errors decrease as the simulation advances towards the final equilibrium state.

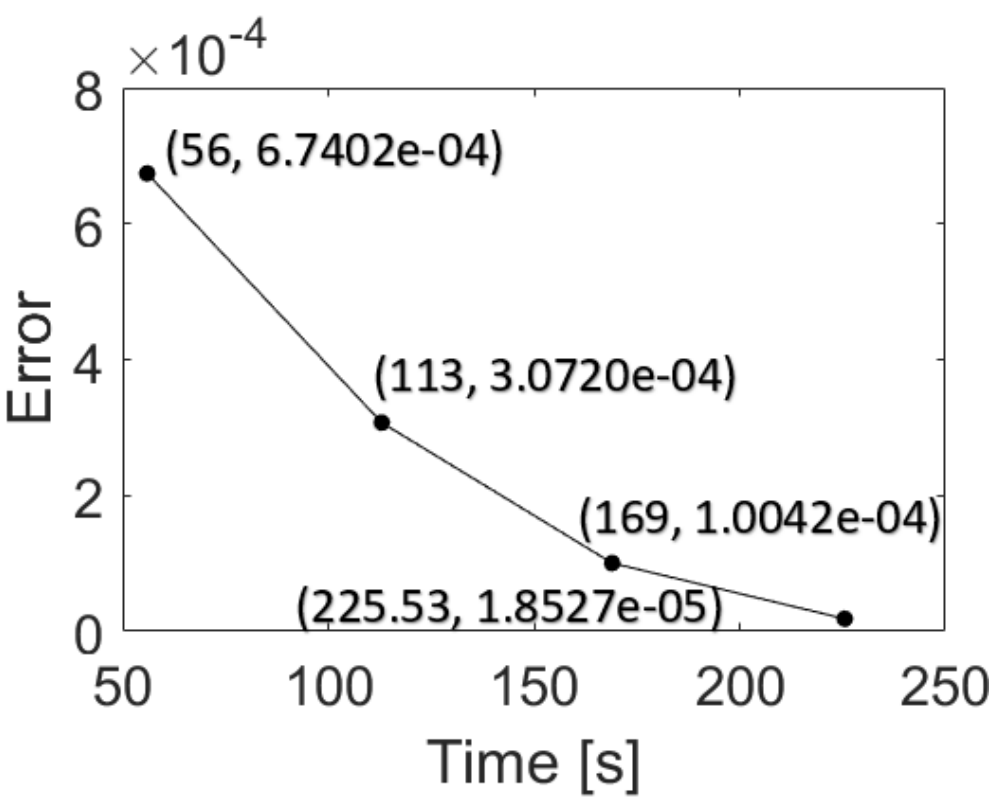

**Figure 7.** Numerical errors of the transient isotropic heat conduction of the proposed FED-FEM compared to ABAQUS; the numerical errors decrease with the increase of time towards equilibrium state.

The orthotropic and anisotropic heat conductions are investigated for the proposed FED-FEM and compared with ABAQUS solutions. Using the same test cube in the isotropic heat conduction, the thermal conductivity $k_{11} = 300$; $k_{22} = 400$; $k_{33} = 200\ W/(m \cdot ℃)$ are used for the orthotropic heat conduction; and $k_{11} = 200$; $k_{12} = 50$; $k_{22} = 300$; $k_{13} = 50$; $k_{23} = 50$; $k_{33} = 400\ \ W/(m \cdot ℃)$ are used for the anisotropic heat conduction. The solutions of the transient orthotropic and anisotropic heat conductions at steady state are presented in Fig. 8, and they are compared with ABAQUS solutions. The steady state for the orthotropic and anisotropic heat conductions are achieved at $t = 225.31\ s$ and $119.32\ s$, respectively.

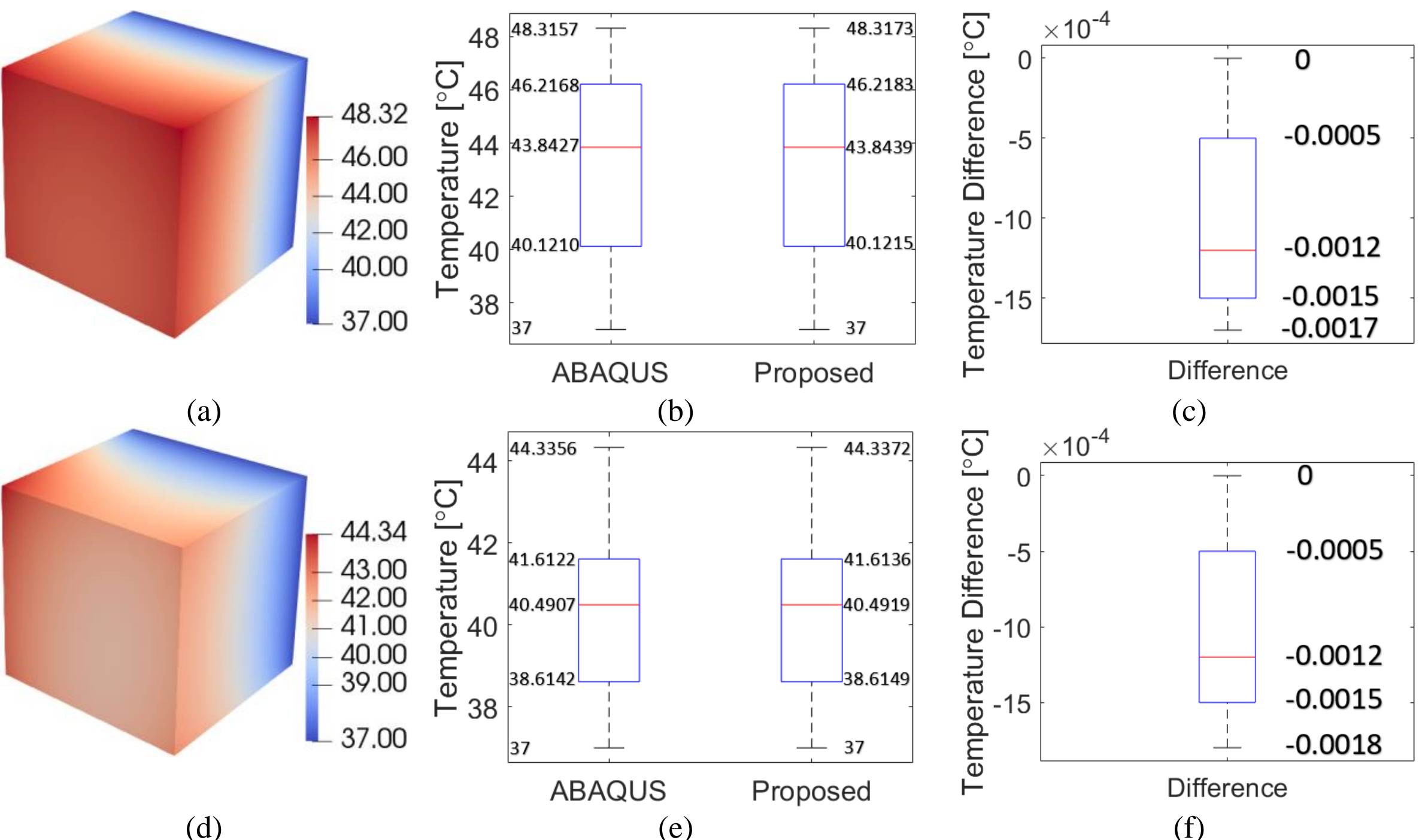

**Figure 8.** Temperature distributions, comparisons of temperature at all nodes, and temperature differences between ABAQUS solutions and proposed FED-FEM at steady state for the case of: (a-c) orthotropic and (d-f) anisotropic heat conductions.

The transient states of the orthotropic and anisotropic heat conductions can be obtained similarly as the transient states of the isotropic heat conduction presented in Fig. 6, and the numerical errors of the transient state of representative time points are presented in Table. 2. Similar to the isotropic heat conduction case, it is also observed that numerical errors in the orthotropic and anisotropic heat conductions decrease as the simulation advances towards the final steady state.

**Table. 2** Numerical errors of the transient orthotropic and anisotropic heat conductions of the proposed FED-FEM compared to ABAQUS.

| Time ($s$) | Error (orthotropic) | Time ($s$) | Error (anisotropic) |
|---|---|---|---|
| 56 | 6.8066e-04 | 30 | 6.9108e-04 |
| 113 | 3.1694e-04 | 60 | 3.2200e-04 |
| 169 | 1.0921e-04 | 90 | 1.0934e-04 |
| 225.31 | 2.6410e-05 | 119.32 | 2.8465e-05 |

4.2.2 Nonlinear temperature-dependent thermal conductivity and specific heat capacity

The proposed FED-FEM is also verified against ABAQUS by considering nonlinear thermal properties, such as the temperature-dependent thermal conductivity and specific heat, for 3-D heat conduction problems. As illustrated in Fig. 9, the tested object is a rectangular plate ($0.2\ m$ x $0.05\ m$ x $0.025\ m$) with three holes in the centre ($\emptyset = 0.025\ m$) separated by a distance $0.05\ m$ each, and it is discretised into a tetrahedral mesh of four-node linear heat transfer tetrahedrons, leading to 13091 elements with 2969 nodes. Material properties are the temperature-dependent isotropic thermal conductivity $k = 200\ W/(m \cdot °\text{C})$ @ $37\ °\text{C}$ and $k = 2000\ W/(m \cdot °\text{C})$ @ $337\ °\text{C}$, medium density $\rho = 1000\ kg/m^3$, specific heat $c = 2000\ J/(kg \cdot °\text{C})$ @ $37\ °\text{C}$ and $c = 8000\ J/(kg \cdot °\text{C})$ @ $337\ °\text{C}$. A linear interpolation between points is employed to determine the temperature-dependent thermal conductivity $k(T)$ and specific heat $c(T)$, i.e., $\Delta k(T) = 6\ W/(m \cdot °\text{C})$ per 1 °C, and $\Delta c(T) = 20\ J/(kg \cdot °\text{C})$ per 1 °C. The initial temperature is $T_0 = 37$ °C. The two side faces and the bottom face of the object are prescribed with a constant temperature $T_r = 37$ °C throughout the simulation, whereas a constant concentrated heat flux $q = 20\ W$ is applied to all nodes at the inner circular faces of the three holes; the adiabatic boundary condition is applied on the remaining faces, and the time step is $\Delta t = 0.005\ s$.

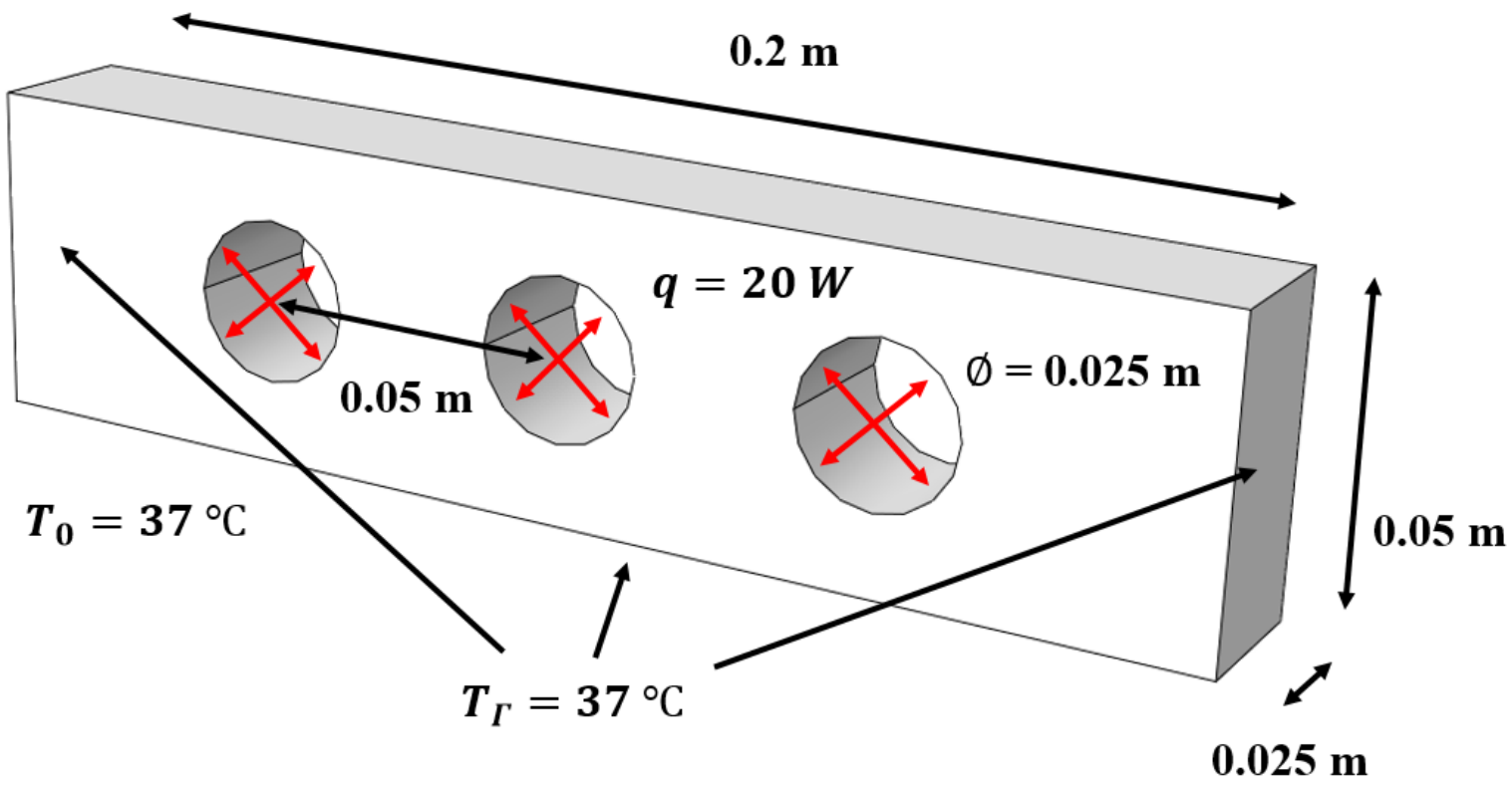

**Figure 9.** Geometry, initial, loading and boundary conditions of the tested plate with three holes: the geometry of the tested object is a rectangular plate ($0.2\ m$ x $0.05\ m$ x $0.025\ m$) with three holes in the centre ($\emptyset = 0.025\ m$) separated by a distance $0.05\ m$ each; the initial temperature is $T_0 = 37$ °C at all nodes; the two side faces and bottom face are prescribed with a constant temperature $T_r = 37$ °C , the inner circular faces of the three holes are prescribed with a constant concentrated heat flux $q = 20\ W$ at all nodes; the adiabatic boundary condition is applied on the remaining faces.

The steady state of the tested plate considering temperature-dependent thermal conductivity and specific heat is achieved at $t = 49.2\ s$, and the transient states at $t = 12.3\ s$, $24.6\ s$ and $36.9\ s$ along with the steady state are compared to the ABAQUS solutions. Due to temperature-dependent thermal properties, the thermal conductivity and specific heat values for nodal calculations are determined by calculating the temperature-dependent thermal property values at element nodes using nodal temperatures and averaging nodal values for the element-level calculation. The errors are 0.0016, 2.4037e-04, 2.4138e-05, and 5.2592e-06 at $t = 12.3\ s$, $24.6\ s$, $36.9\ s$ and $49.2\ s$, respectively. It can be seen from Fig. 10 that there is good agreement with those of the proposed FED-FEM compared to ABAQUS with the same set of plate mesh and simulation parameters.

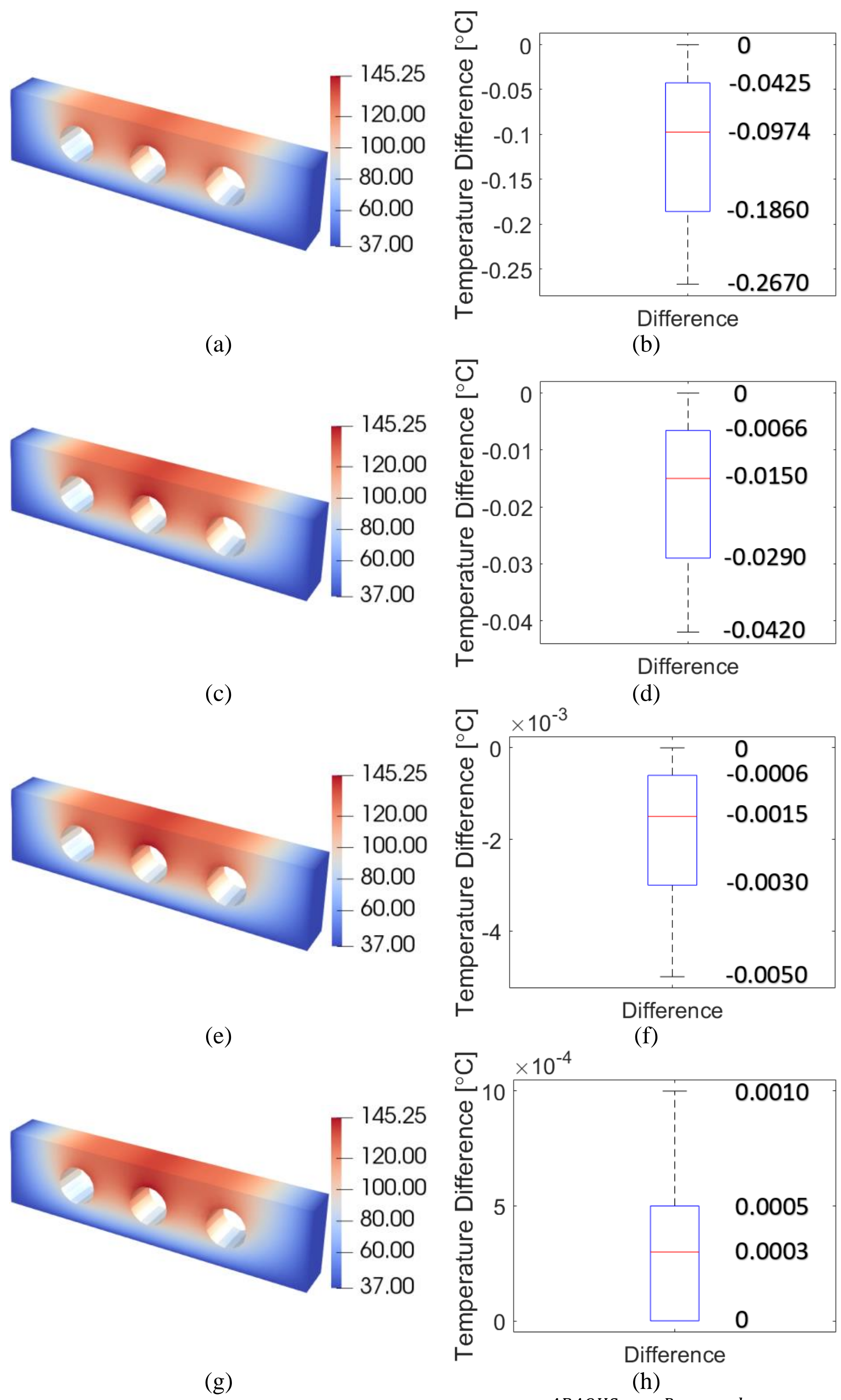

**Figure 10.** Temperature distributions and temperature differences ($T^{ABAQUS} - T^{Proposed}$) of the transient/steady states of the tested plate at (a-b) $t = 12.3\ s$; (c-d) $t = 24.6\ s$; (e-f) $t = 36.9\ s$; and (g-h) $t = 49.2\ s$ (equilibrium).

### 4.3 Heat convection

The proposed FED-FEM is also verified against ABAQUS for 3-D heat convection problems. As illustrated in Fig. 11, the tested object is a rectangular air fin cooler (0.09 $m$ x 0.06 $m$ x 0.08 $m$) with five vertical fins separated by four gaps, the dimension of each gap is 0.01 $m$ x 0.05 $m$ x 0.08 $m$. The air fin cooler is discretised into a tetrahedral mesh of four-node linear heat transfer tetrahedrons, leading to 70028 elements with 15287 nodes. Material properties are the isotropic thermal conductivity $k = 200\ W/(m \cdot °C)$, medium density $\rho = 3000\ kg/m^3$, and specific heat $c = 200\ J/(kg \cdot °C)$. The initial temperature is $T_0 = 37$ °C. The bottom face of the air fin cooler is prescribed with a constant temperature $T_\Gamma = 37$ °C throughout the simulation. Heat convection is applied on all the remaining faces with a convection heat transfer coefficient $h = 200\ W/(m^2 \cdot °C)$ and an ambient temperature $T_a = 0$ °C. In the ABAQUS simulation, a concentrated film condition is applied on the heat convection surfaces with an associated nodal area of $8.655 \times 10^{-6}\ m^2$, which is determined by calculating the sum of convection surface areas and dividing by the number of nodes on the convection surfaces. The time step is $\Delta t = 0.001\ s$.

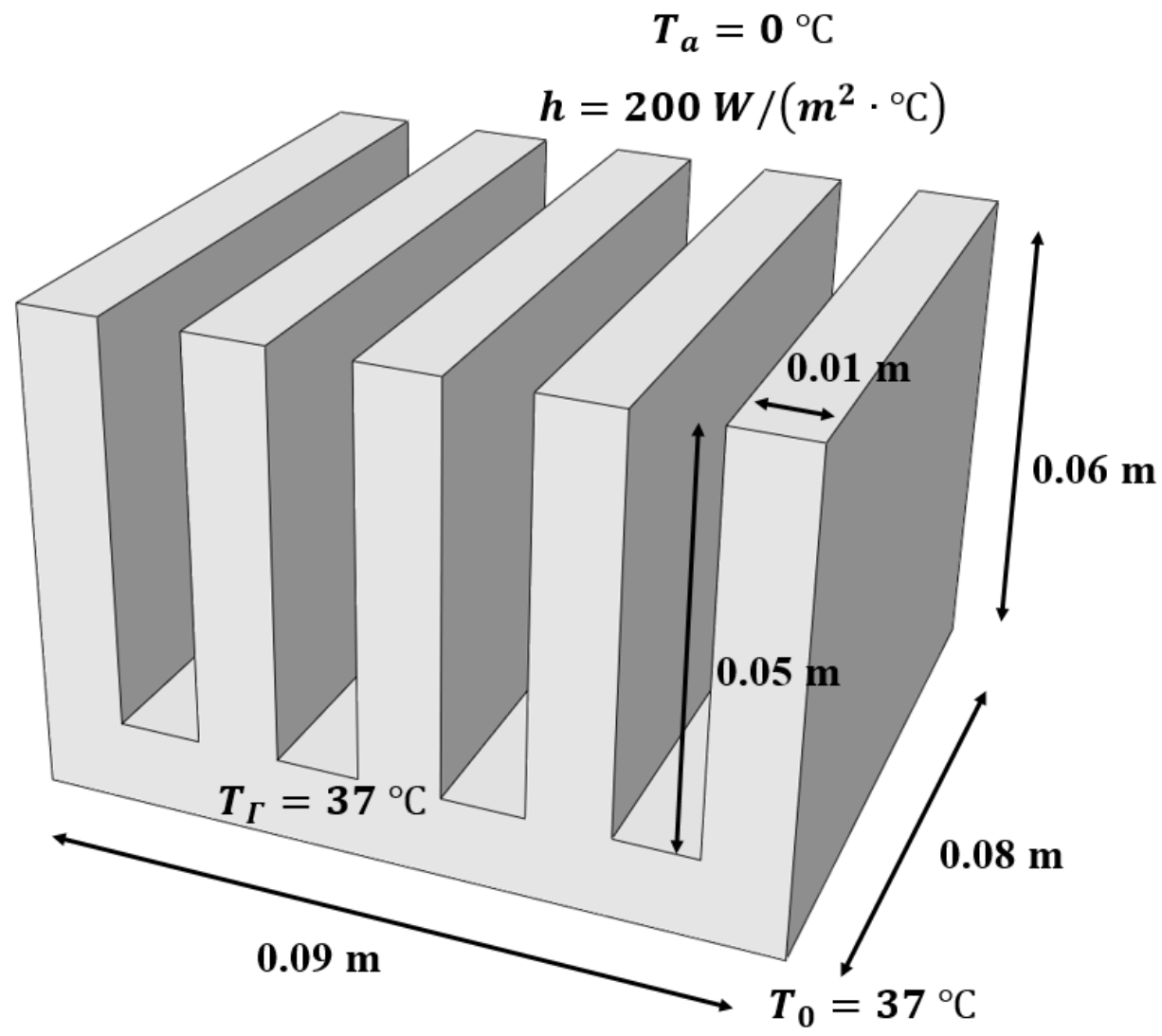

**Figure 11.** Geometry, initial, and boundary conditions of the tested air fin cooler: the geometry of the tested object is a rectangular air fin cooler ($0.09\ m$ x $0.06\ m$ x $0.08\ m$) with five vertical fins separated by four gaps, the dimension of each gap is $0.01\ m$ x $0.05\ m$ x $0.08\ m$; the initial temperature is $T_0 = 37$ °C at all nodes; the bottom face is prescribed with a constant temperature $T_\Gamma = 37$ °C; heat convection is applied on all the remaining faces with a convection heat transfer coefficient $h = 200\ W/(m^2 \cdot °C)$ and an ambient temperature $T_a = 0$ °C.

The steady state of the tested air fin cooler is achieved at $t = 34.6\ s$, and the transient states at $t = 8.6\ s$, $17.3\ s$ and $26.0\ s$ along with the steady state are compared to the ABAQUS solutions. The errors are 0.0032, 3.6918e-04, 6.6706e-05, and 1.2529e-05 at $t = 8.6\ s$, $17.3\ s$, $26.0\ s$ and $34.6\ s$, respectively. It can be seen from Fig. 12 that there is good agreement with those of the FED-FEM compared to ABAQUS with the same set of mesh and simulation parameters.

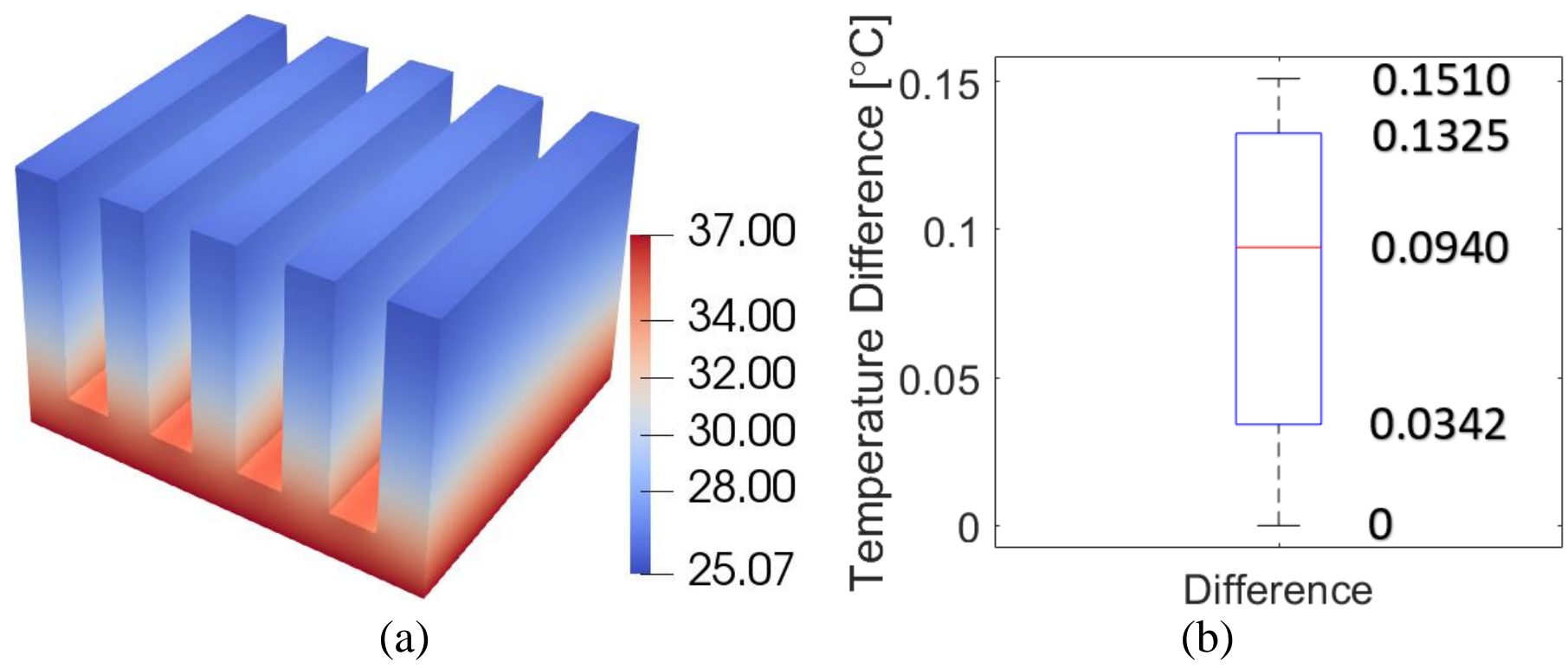

(a) (b)

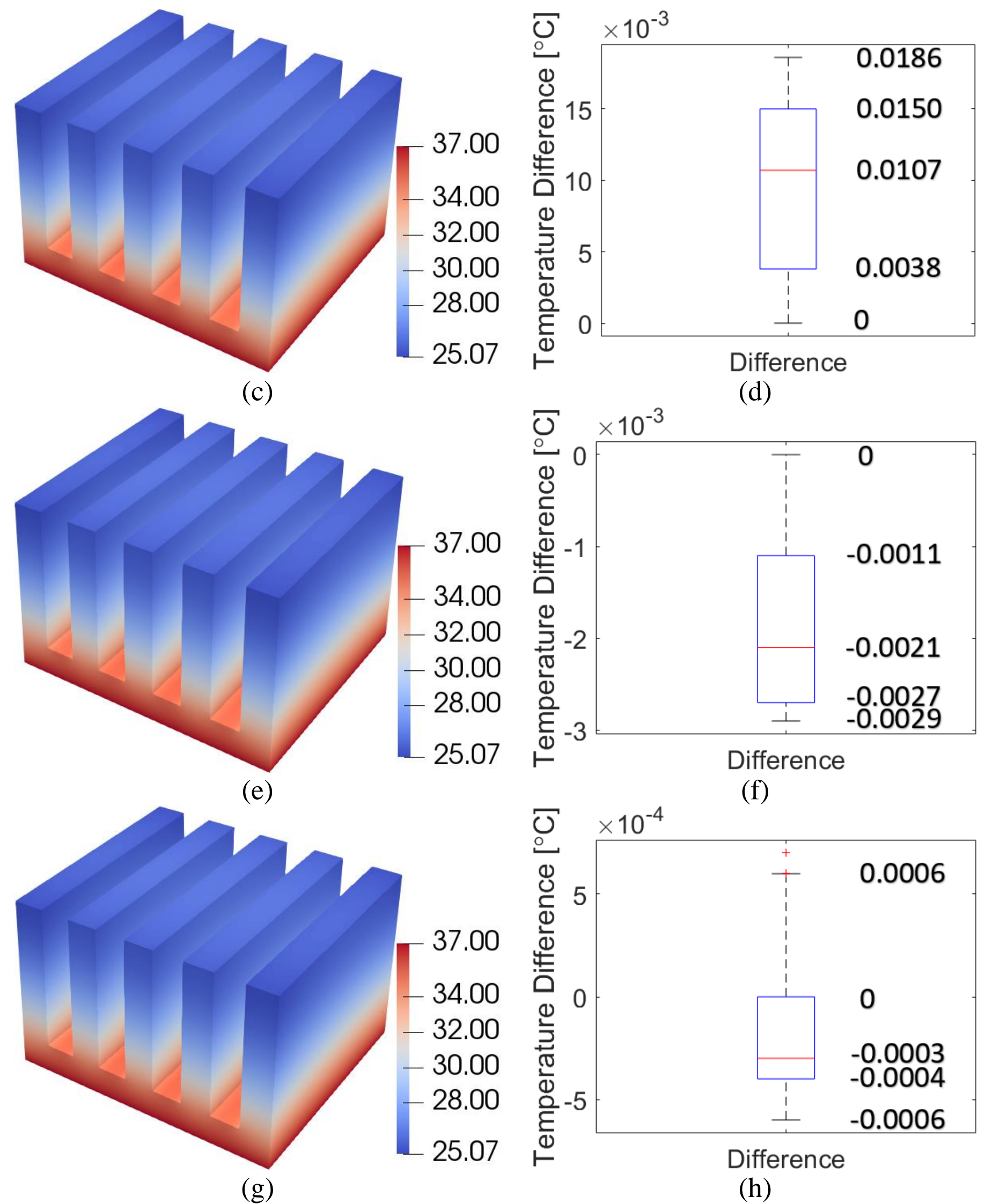

**Figure 12.** Temperature distributions and temperature differences ($T^{ABAQUS} - T^{Proposed}$) of the transient/steady states of the tested air fin cooler at (a-b) $t = 8.6\ s$; (c-d) $t = 17.3\ s$; (e-f) $t = 26.0\ s$; and (g-h) $t = 34.6\ s$ (equilibrium).

### 4.4 Heat radiation

The proposed FED-FEM is also verified against ABAQUS for 3-D heat radiation problems. As illustrated in Fig. 13, the tested object is a U-shaped radiator ($0.2\ m$ x $0.08\ m$ x $0.1\ m$) with a circular cut in the centre ($\emptyset = 0.12\ m$). The U-shaped radiator is discretised into a tetrahedral mesh of four-node linear heat transfer tetrahedrons, leading to 8307 elements with 1843 nodes. Material properties are the isotropic thermal conductivity $k = 200\ W/(m \cdot$ ℃), medium density $\rho = 2000\ kg/m^3$, and specific heat $c = 300\ J/(kg \cdot$ ℃). The initial temperature is $T_0 = 37$ ℃ at all nodes. The bottom face is prescribed with a constant temperature $T_\Gamma = 37$ ℃ throughout the simulation. Heat radiation is exerted on all the remaining faces with the Stefan-Boltzmann constant $\sigma = 5.67 \times 10^{-8}\ Wm^{-2}K^{-4}$, emissivity $\varepsilon = 0.66$, absolute zero temperature $T_z = -273.15$ ℃, and an ambient temperature $T_a = -50$ ℃. In the ABAQUS simulation, a concentrated radiation to ambient condition is applied on the heat radiation faces with an associated nodal area of $3.882 \times 10^{-5}\ m^2$, which is determined by calculating the sum of areas of radiation faces and dividing by the number of nodes on the radiation faces. The time step is $\Delta t = 0.001\ s$.

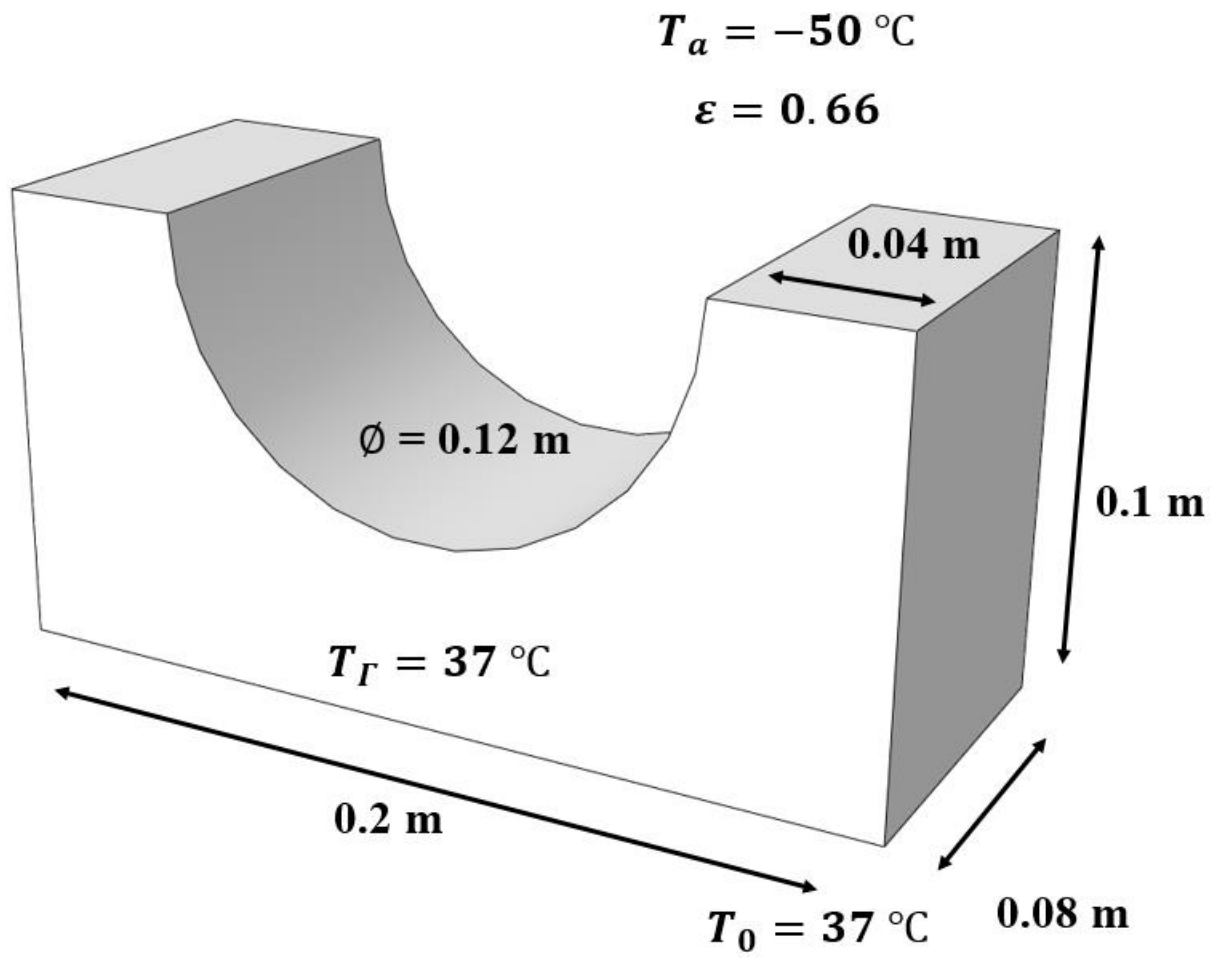

**Figure 13.** Geometry, initial, and boundary conditions of the tested U-Shaped radiator: the geometry of the tested object is a U-Shaped radiator ($0.2\ m$ x $0.08\ m$ x $0.1\ m$) with a circular cut ($\emptyset = 0.12\ m$) in the centre; the initial temperature is $T_0 = 37$ ℃ at all nodes; the bottom face is prescribed with a constant temperature $T_\Gamma = 37$ ℃; heat radiation is applied on all the remaining faces with the Stefan-Boltzmann constant $\sigma = 5.67 \times 10^{-8}\ Wm^{-2}K^{-4}$, emissivity $\varepsilon = 0.66$, absolute zero temperature $T_z = -273.15$ ℃, and an ambient temperature $T_a = -50$ ℃.

The steady state of the tested U-shaped radiator is achieved at $t = 46.23\ s$, and the transient states at $t = 11.56\ s$, $23.12\ s$ and $34.68\ s$ along with the steady state are compared to the ABAQUS solutions. The errors are 2.3482e-04, 1.9701e-04, 6.9417e-05, and 4.0355e-05 at $t = 11.56\ s$, $23.12\ s$, $34.68\ s$ and $46.23\ s$, respectively. It can be seen from Fig. 14 that there is good agreement with those of the proposed FED-FEM compared to ABAQUS with the same set of mesh and simulation parameters.

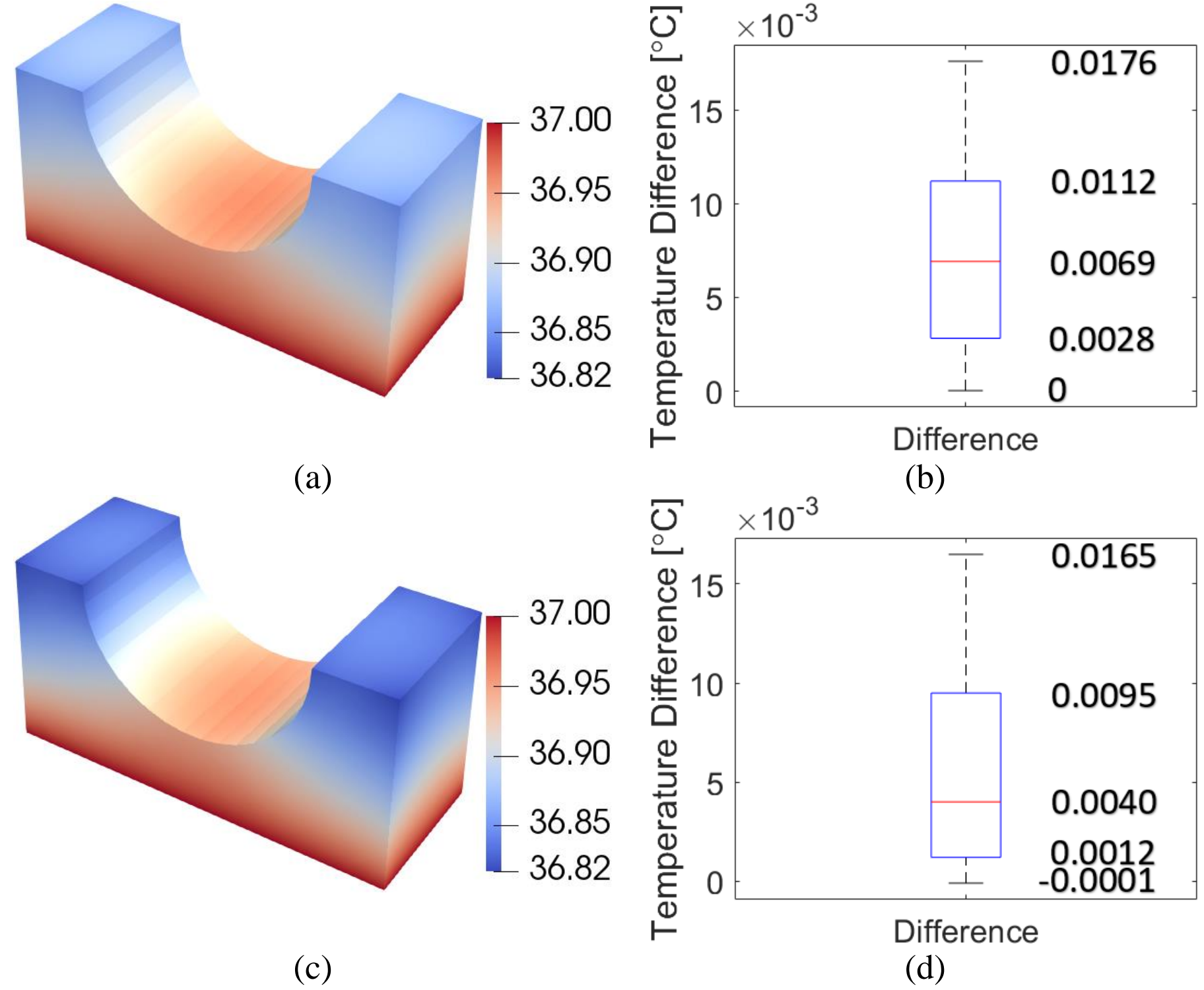

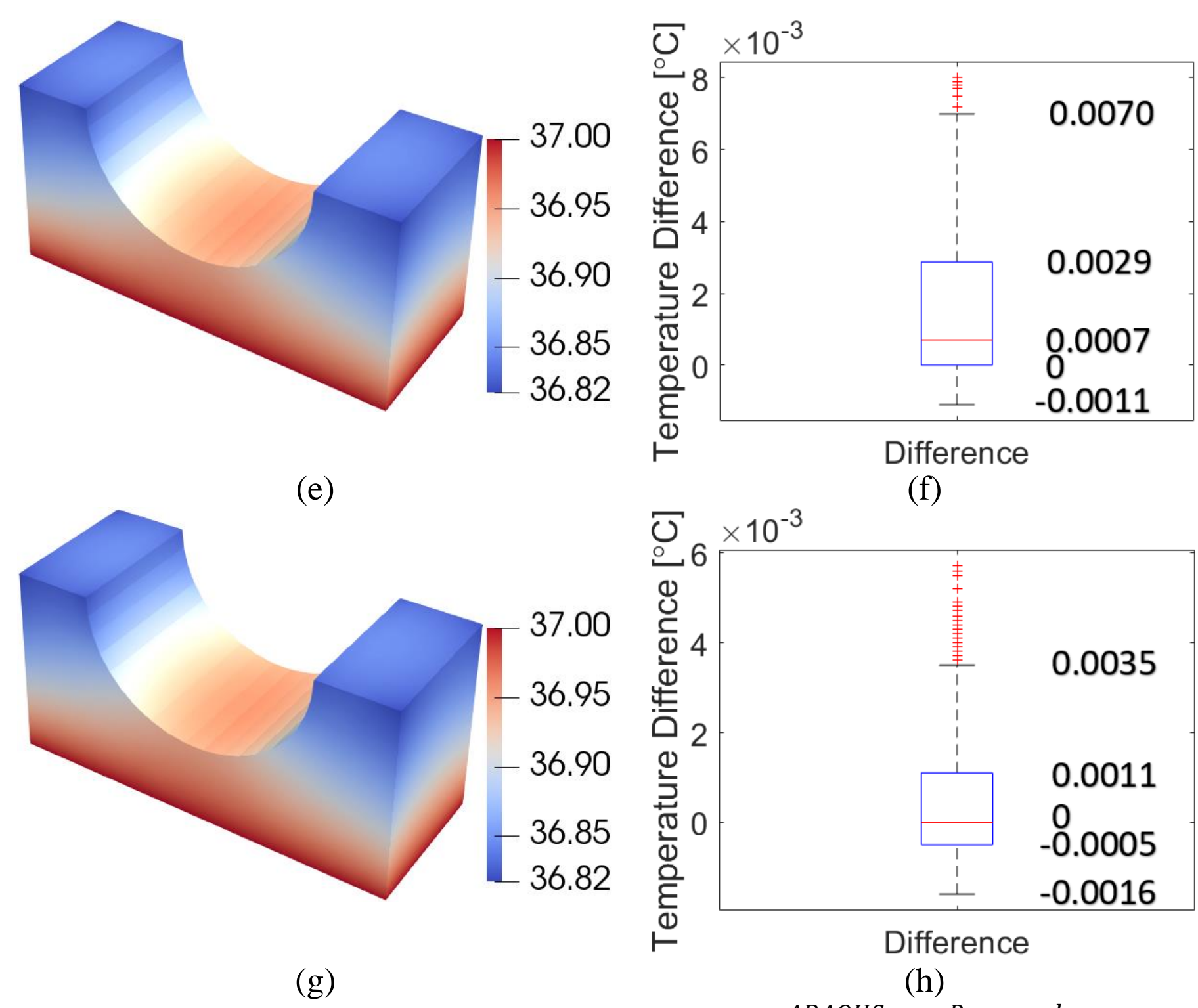

**Figure 14.** Temperature distributions and temperature differences ($T^{ABAQUS} - T^{Proposed}$) of the transient/steady states of the tested U-shaped radiator at (a-b) $t = 11.56\ s$; (c-d) $t = 23.12\ s$; (e-f) $t = 34.68\ s$; and (g-h) $t = 46.23\ s$ (equilibrium).

### 4.5 Concentrated heat source

Highly concentrated thermal energy can lead to a steep local temperature gradient in the vicinity of the heated medium [37]. For numerical accuracy, very fine meshes are usually used near the thermal gradient concentration zone, while other zones use a coarse mesh. This arrangement may lead to distorted transition elements, especially when using tetrahedral elements, leading to a degradation of solution precisions. To verify the accuracy of the proposed FED-FEM in the event of high thermal gradient concentration, a concentrated heat source $q = 10\ W$ is applied to a circular zone ($\emptyset = 0.006\ m$) on the surface of the 3-D rectangular plate illustrated in Fig. 15. The rectangular plate ($0.2\ m$ x $0.03\ m$ x $0.05\ m$) is discretised into a tetrahedral mesh of four-node linear heat transfer tetrahedrons, leading to 2574 elements with 619 nodes. Material properties are the isotropic thermal conductivity $k = 400\ W/(m \cdot ℃)$, medium density $\rho = 2000\ kg/m^3$, and specific heat $c = 300\ J/(kg \cdot ℃)$. The initial temperature is $T_0 = 37$ ℃ at all nodes. The back face is prescribed with a constant temperature $T_\Gamma = 37$ ℃ throughout the simulation, whereas the adiabatic boundary condition is applied on the remaining faces. The time step is $\Delta t = 0.0001\ s$.

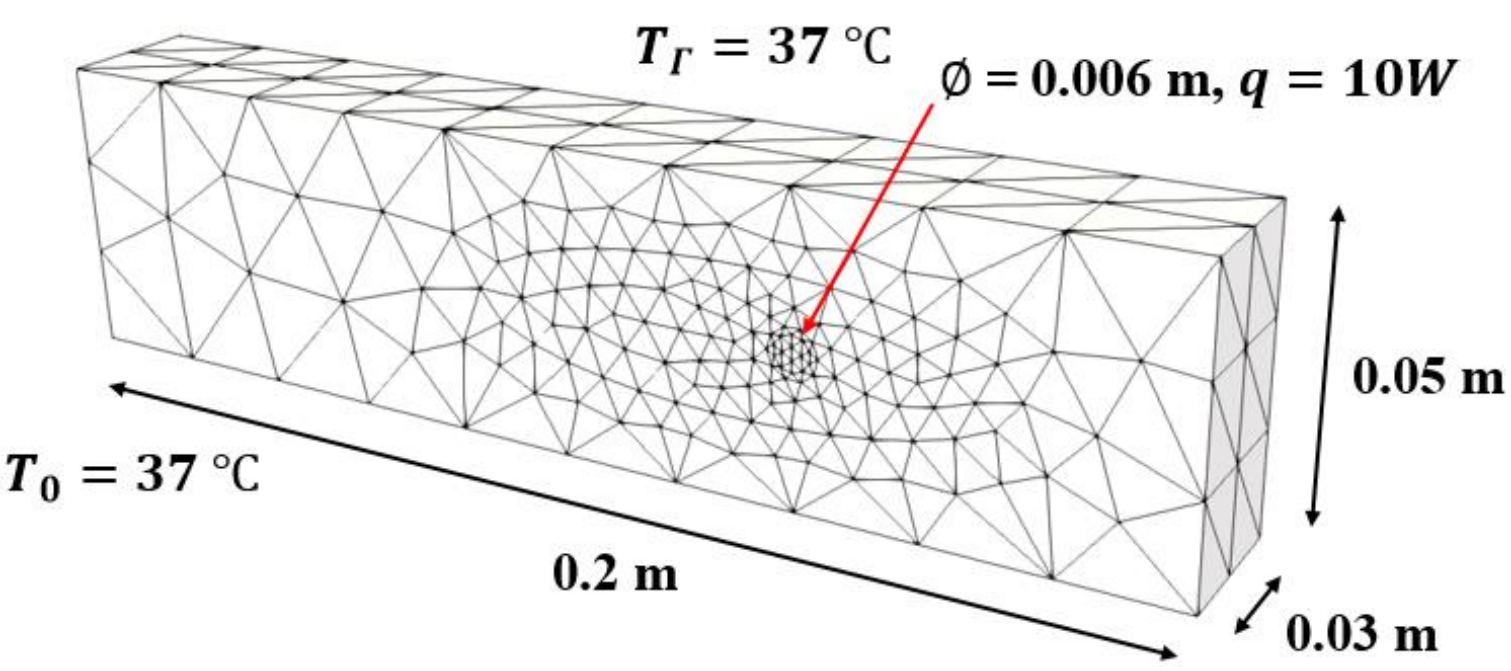

**Figure 15.** Geometry, initial, and boundary conditions of the tested rectangular plate ($0.2\ m$ x $0.03\ m$ x $0.05\ m$) subject to a concentrated heat source $q = 10\ W$ applied to the circular zone ($\emptyset = 0.006\ m$) on the surface of the plate; the initial temperature is $T_0 = 37$ ℃ at all nodes; the back face is prescribed with a constant temperature $T_\Gamma = 37$ ℃; and the adiabatic boundary condition is applied on the remaining faces.

The steady state of the tested rectangular plate is achieved at $t = 3.47\ s$, and the transient states at $t = 0.87\ s$, $1.74\ s$ and $2.61\ s$ along with the steady state are compared to the ABAQUS solutions. The errors are 1.4678e-04, 1.4534e-04, 1.0776e-04, and 2.4387e-05 at $t = 0.87\ s$, $1.74\ s$, $2.61\ s$ and $3.47\ s$, respectively. It is observed from Fig. 16 that the proposed FED-FEM can achieve numerical accuracy comparable to those of the ABAQUS. This shows that the proposed FED-FEM algorithm can be used as an alternative method to conventional FEM for thermal problems.

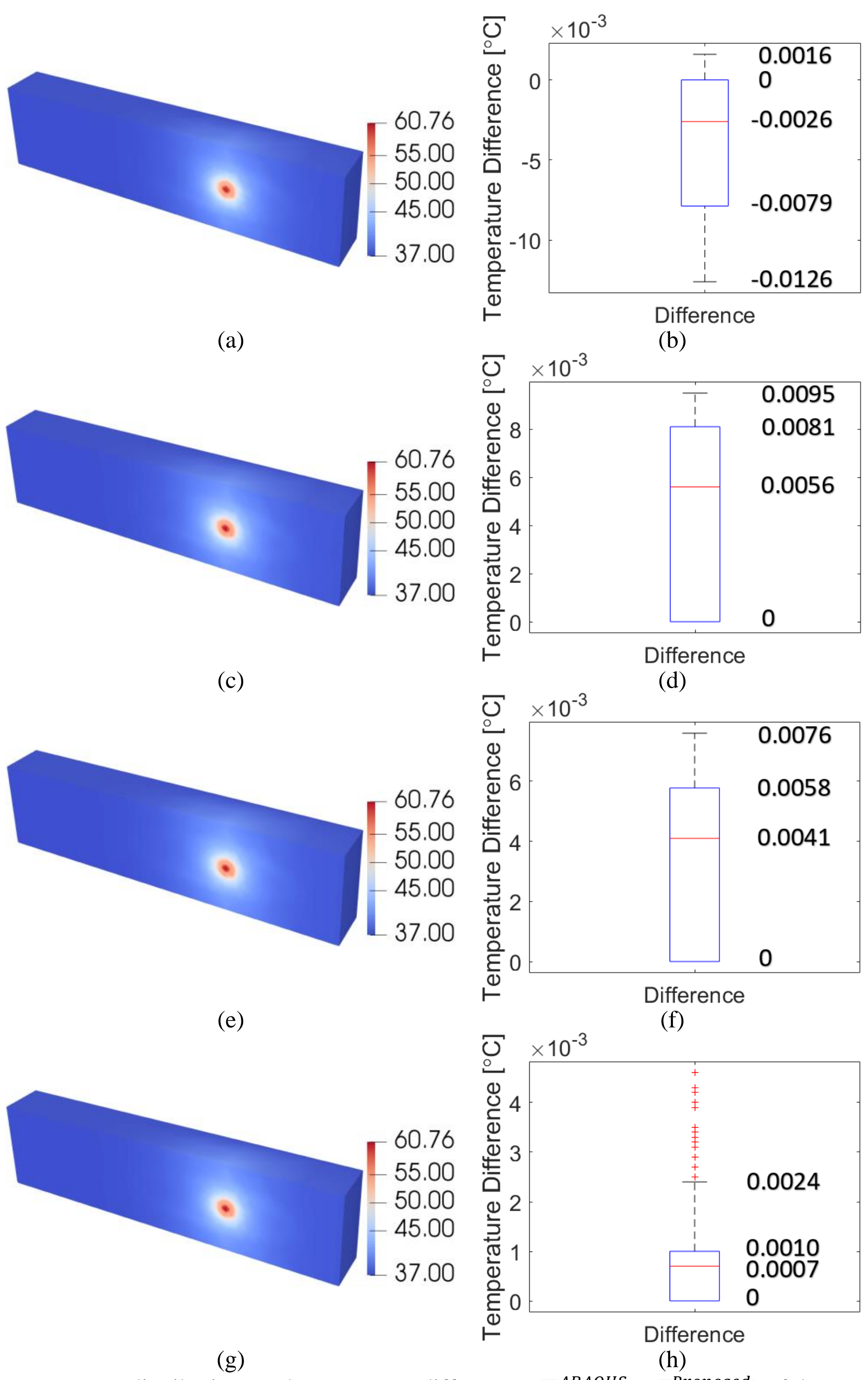

**Figure 16.** Temperature distributions and temperature differences ($T^{ABAQUS} - T^{Proposed}$) of the transient/steady states of the tested rectangular plate at (a-b) $t = 0.87\ s$; (c-d) $t = 1.74\ s$; (e-f) $t = 2.61\ s$; and (g-h) $t = 3.47\ s$ (equilibrium).

### 4.6 Computational performance

The proposed FED-FEM is implemented in C++ and evaluated on an Intel(R) Core(TM) i7-6700 CPU @ 3.40 GHz and 32.0 GB RAM PC. A cube model is used to evaluate the computation time, and results are compared to the ABAQUS solution times under same conditions. The cube is discretised into six mesh densities which are the (1445 nodes, 6738 elements), (1840 nodes, 8724 elements), (2644 nodes, 12852 elements), (3339 nodes, 16371 elements), (5304 nodes, 26550 elements), and (7872 nodes, 40021 elements). Both the proposed FED-FEM and ABAQUS are executed using a single-thread central processing unit (CPU) process without multi-thread CPU nor parallel graphics processing unit (GPU) acceleration. The simulation is performed at $\Delta t = 0.005\ s$ for 2000 time steps for a simulation of $t = 10.0\ s$ transient heat transfer process. Fig. 17 illustrates a comparison of computation times of isotropic, orthotropic, and anisotropic cases using the proposed FED-FEM considering temperature-independent thermal properties. As mentioned in Section 3.3, since the thermal conductivity and the associated thermal stiffness matrix may be pre-computed for the linear heat transfer problems, the three cases rendered nearly identical computation times. The proposed methodology is able to perform a calculation at $t = 2.376\ ms$ per time step at (7872 nodes, 40021 elements) and completes the simulation (2000 time steps) in $4752\ ms$. Furthermore, since the computation time $t = 2.376\ ms$ per time step is lower than the physical time step size $\Delta t = 0.005\ s = 5\ ms$, the proposed FED-FEM is able to provide real-time computational performance for the cube model (7872 nodes, 40021 tetrahedrons) with the mentioned hardware settings.

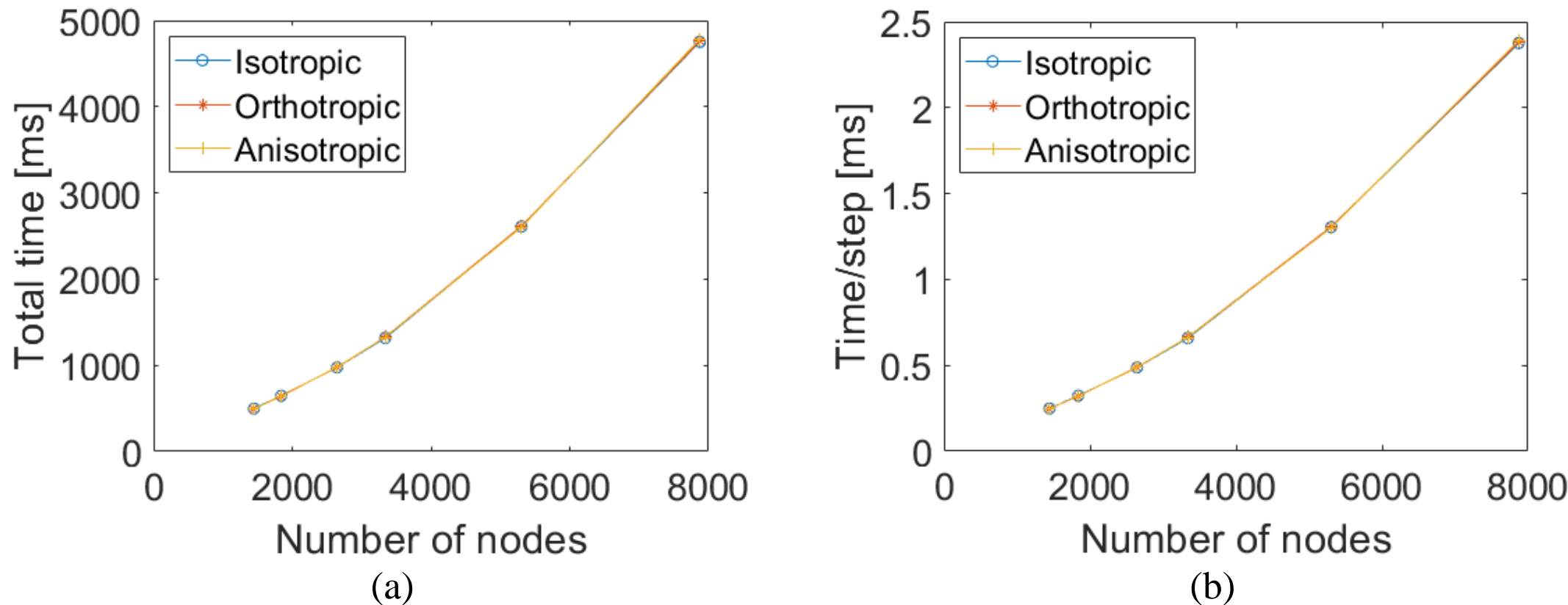

(a) (b)

**Figure 17.** Comparison of computation time of isotropic, orthotropic and anisotropic cases of the cube model for different mesh sizes using the proposed FED-FEM considering temperature-independent thermal properties: (a) total computation time, and (b) computation time per time step.

Fig. 18 illustrates a comparison of computation times between the proposed FED-FEM and ABAQUS for the isotropic case utilising different mesh sizes. At (7872 nodes, 40021 elements), the proposed FED-FEM consumes $t = 4752\ ms$ to complete the simulation with a computation time $t = 2.376\ ms$ per step; however, the ABAQUS takes $t = 1665.5\ s = 1.6655 \times 10^{6}\ ms$ with a computation time $t = 832.75\ ms$ per step; this leads to a computation time reduction of 350.48 times using the proposed FED-FEM over the ABAQUS solution method.

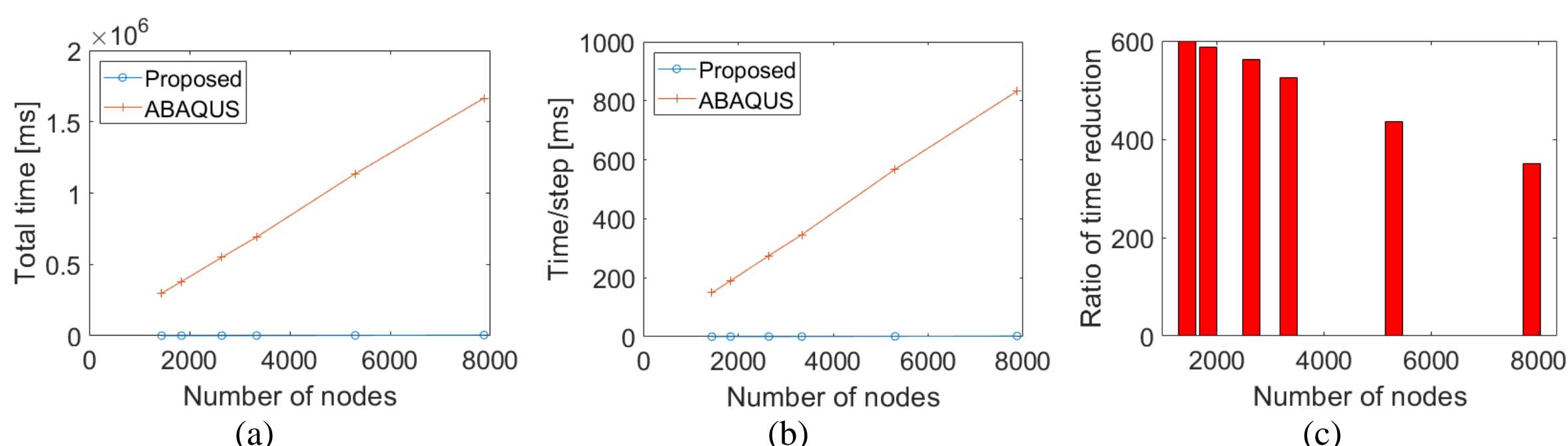

(a) (b) (c)

**Figure 18.** Comparison of computation times between the proposed FED-FEM and ABAQUS for the isotropic case considering temperature-independent thermal properties in terms of (a) total computation time, (b) computation time per time step, and (c) ratio of computation time reduction ($t_{Abaqus}/t_{Proposed}$).

Furthermore, the computation time of the proposed methodology is evaluated for the temperature-dependent (TD) thermal properties case. As mentioned in Sections 3.1 and 3.2, compared to the temperature-independent (TI) thermal properties case, the thermal mass matrix and thermal stiffness matrix require an update at each time step for the TD case due to the temperature-dependent specific heat and thermal conductivity. However, pre-computation of the coefficient matrix **A** and matrix $\mathbf{G}(\mathbf{x})$ can still be evaluated before the commencement of the time-stepping procedure to improve computational efficiency. Fig. 19 illustrates a comparison between the TI and TD cases for the isotropic cube model. The TD case consumes more time than the TI case, completing the simulation (2000 time steps, 7872 nodes, 40021 elements) in $t = 9312\ ms$ with an average computation time $t = 4.656\ ms$ per time step compared to the TI case $t = 4752\ ms$ and $t = 2.376\ ms$, respectively. As seen from Fig. 19(c), the TD case consumes an approximately 2-2.5 times more computation time than the TI case. Using a linear interpolation between points, the real-time computational performance of the TD case is estimated to be obtained at (8360 nodes, 42600 elements, $t = 5\ ms$ per step) for the cube model with the mentioned hardware settings.

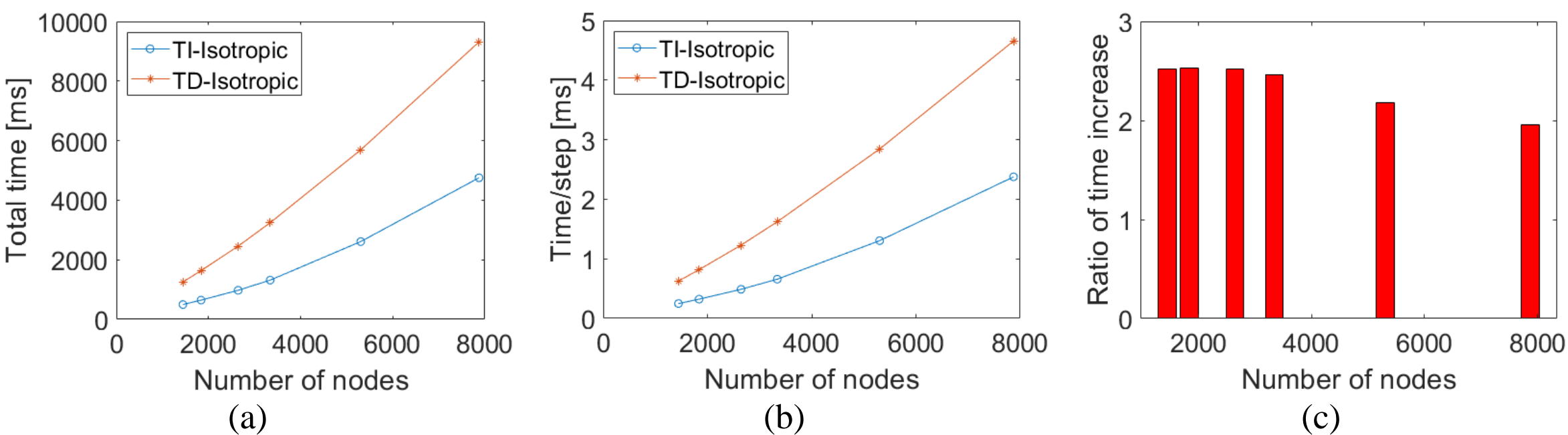

**Figure 19.** Comparison of computation times between the temperature-independent (TI) and temperature-dependent (TD) thermal properties using the proposed FED-FEM for the isotropic case of the cube model in terms of (a) total computation time, (b) computation time per time step, and (c) ratio of computation time increase ($t_{TD}/t_{TI}$).

Fig. 20 presents a comparison of computation times between the proposed FED-FEM and ABAQUS for the TD-isotropic case utilising different mesh sizes. The solution times of both proposed methodology and ABAQUS are increased due to the TD thermal properties. At (7872 nodes, 40021 elements), the proposed FED-FEM consumes $t = 9312\ ms$ to complete the simulation with a computation time $t = 4.656\ ms$ per step; however, the ABAQUS takes $t = 1682.3\ s = 1.6823 \times 10^6\ ms$ with a computation time $t = 841.15\ ms$ per step; this leads to a computation time reduction of 180.66 times using the proposed FED-FEM over the ABAQUS solution method.

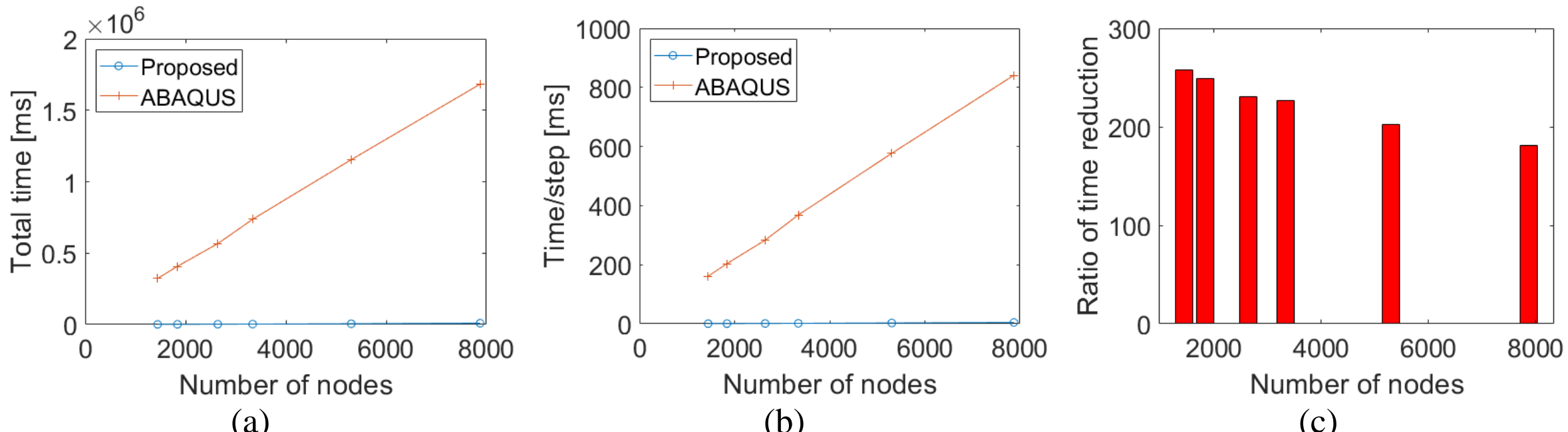

**Figure 20.** Comparison of computation times between the proposed FED-FEM and ABAQUS for the TD-isotropic case in terms of (a) total computation time, (b) computation time per time step, and (c) ratio of computation time reduction ($t_{Abaqus}/t_{Proposed}$).

## 5. Discussion

The proposed FED-FEM provides a fast and effective means for modelling of transient heat transfer problems. It achieves fast computation of transient heat transfer problems via the (i) explicit time integration, (ii) element-wise thermal load computation, (iii) explicit formulation of nodal temperature calculation, (iv) pre-computation of simulation parameters, and (v) computationally efficient finite elements. With the proposed fast explicit formulation, a number of simulation parameters can be pre-computed offline prior to the online simulation, promoting the efficiency of online computation. Using the mentioned hardware settings, the proposed FED-FEM achieves a computation time reduction of 350.48 times at (7872 nodes, 40021 tetrahedrons) considering temperature-independent thermal properties and 180.66 times considering temperature-dependent thermal properties compared to the ABAQUS solution times under same conditions, far exceeding the computational performance of the commercial FEM codes. With the simulation time step size of $\Delta t = 0.005\ s = 5\ ms$, the real-time computational performance is achieved for the TI case at (7872 nodes, 40021 tetrahedrons, $t = 2.376\ ms$ per step) using the proposed FED-FEM, and it is estimated to simulate (14164 nodes, 73025 tetrahedrons) in real-time at $t = 5\ ms$ per step using a linear interpolation. The real-time computational performance of the TD case is estimated to be obtained at (8360 nodes, 42600 elements, $t = 5\ ms$ per step) for the same cube model. Owing to the explicit formulation of nodal temperature calculation, the proposed FED-FEM is well suited for multi-thread CPU implementation and GPU parallel implementation, where individual nodal unknowns may be computed independently using independent equations, for even faster solution time of transient heat transfer problems. Thus, the proposed FED-FEM algorithm constitutes a step towards a real-time engineering application. However, it needs to be noted that precautions need to be taken for the chosen simulation time step size due to the conditional stability of the explicit time integration. The stability analysis of the explicit integration is presented in Section 3.4, and readers may refer to [31] for further detailed discussions.

While achieving the fast solution time to the transient heat transfer problems, the proposed FED-FEM also passes the standard patch tests for numerical convergence and maintains comparable solution accuracy to ABAQUS. It is observed that a typical 1.0e-5 level of accuracy can be maintained at the equilibrium state using the proposed methodology while a less accurate solution is observed along the path. This is mainly due to the use of mass lumping for the mass and specific heat for the thermal mass matrix. It is known that the final steady-state solutions of the transient heat transfer problem are identical regardless of the locations of integration points in the thermal mass matrix, but the modification of integration points in the mass may lead to a sacrifice of numerical accuracy along the path [15]. However, a typical 1.0e-4 level of accuracy can still be maintained by the proposed FED-FEM in the simulation of transient states of heat transfer. Nonlinear thermal properties such as the temperature-dependent thermal conductivity and specific heat and nonlinear boundary conditions such as the heat convection and radiation can be handled effectively at the element-level computation and explicit nodal temperature formulation, respectively. The results from these finite element experiments are promising, demonstrating that a high level of precision can be achieved in real-time simulations using the proposed methodology. Some newly developed numerical methods such as the node-based smoothed FEM [24, 38], smoothed point interpolation method [39], gradient weighted FEM [23], and "Tri-Prism" FEM [21] are focused more on the numerical accuracy, convergence and stability rather than the real-time computational performance, the proposed FED-FEM algorithm may be incorporated into these methods for improved computational performance.

## 6. Conclusions

This paper presents a fast explicit dynamics finite element algorithm for transient heat transfer problems, and it is well suited for real-time applications. The proposed FED-FEM can lead to fast computation due to (i) formulating the solution procedure based on the explicit time integration to efficiently integrate the dynamic heat transfer equation in the temporal domain, (ii) devising an element-level computation to obtain nodal thermal loads, eliminating the need for assembling the global thermal stiffness matrix, and subsequently leading to (iii) an explicit formulation for nodal temperature calculation, eliminating the need for iterations anywhere in the algorithm, (iv) pre-computing the constant matrices and simulation parameters offline to facilitate the online computational performance, and (v) utilising computationally efficient finite elements for efficient thermal response calculation in the spatial domain. Nonlinear thermal material properties and nonlinear boundary conditions can be accommodated by the proposed methodology. Simulations and comparison analyses demonstrate that not only can the proposed FED-FEM achieve fast solution times but also maintain numerical accuracy and convergence compared with those of the commercial finite element analysis packages. The computation time of the proposed methodology is significantly lower than that of standard FEM, capable of achieving real-time computational performance for real-time simulation of transient heat transfer problems.

Future research work will focus on improvement and extension of the proposed FED-FEM. The proposed methodology will be extended for integration with GPU hardware to further facilitate the computational performance. The current implementation only utilises a single-thread computing power of CPU without efforts on hardware acceleration. It is expected that the GPU-accelerated solution method will significantly improve the computational performance of the proposed FED-FEM. In addition, the proposed methodology will also be extended for integration into real-time-based applications, such as the virtual-reality-based or augmented-reality-based applications. Novel algorithms will be developed in the future to predict the temperature field in real-time for efficient and accurate prediction and control of temperature.

**Acknowledgment**

This work is funded by the Australian National Health and Medical Research Council (NHMRC) Grant APP1093314.

**References**

[1] Karaki W, Rahul, Lopez CA, Borca-Tasciuc DA, De S. A continuum thermomechanical model of in vivo electrosurgical heating of hydrated soft biological tissues. International Journal of Heat and Mass Transfer. 2018;127:961-74.
[2] Feng SZ, Han X, Wang G. An efficient on-line algorithm for the optimal design of multi-material structures under thermal loads. International Journal of Thermal Sciences. 2018;132:567-77.
[3] Li XG, Yi LP, Yang ZZ, Chen YT, Sun J. Coupling model for calculation of transient temperature and pressure during coiled tubing drilling with supercritical carbon dioxide. International Journal of Heat and Mass Transfer. 2018;125:400-12.
[4] Wilson EL, Bathe KJ, Peterson FE. Finite-Element Analysis of Linear and Nonlinear Heat Transfer. Nuclear Engineering and Design. 1974;29:110-24.
[5] Zienkiewicz O, Taylor R, Zhu J. The finite element method: its basis and fundamentals: Butterworth-Heinemann; 2005.
[6] Duda P. A general method for solving transient multidimensional inverse heat transfer problems. International Journal of Heat and Mass Transfer. 2016;93:665-73.
[7] Shen B, Shih AJ, Xiao GX. A Heat Transfer Model Based on Finite Difference Method for Grinding. Journal of Manufacturing Science and Engineering-Transactions of the Asme. 2011;133:031001--10.
[8] Chow JH, Zhong ZW, Lin W, Khoo LP. The development of a simple semi-empirical method to obtain a predictive model of the convective heat loss of a solid surface of practical machine size by finite difference and experimentation. Applied Thermal Engineering. 2015;75:789-99.
[9] Sun YJ, Zhang XB. Analysis of transient conduction and radiation problems using lattice Boltzmann and finite volume methods. International Journal of Heat and Mass Transfer. 2016;97:611-7.
[10] Sun YJ, Zhang XB. A hybrid strategy of lattice Boltzmann method and finite volume method for combined conduction and radiation in irregular geometry. International Journal of Heat and Mass Transfer. 2018;121:1039-54.
[11] Wen J, Khonsari MM. Transient heat conduction in rolling/sliding components by a dual reciprocity boundary element method. International Journal of Heat and Mass Transfer. 2009;52:1600-7.
[12] Simoes N, Tadeu A, Antonio J, Mansur W. Transient heat conduction under nonzero initial conditions: A solution using the boundary element method in the frequency domain. Engineering Analysis with Boundary Elements. 2012;36:562-7.
[13] Xiao YH, Zhan HF, Gu YT, Li QH. Modeling heat transfer during friction stir welding using a meshless particle method. International Journal of Heat and Mass Transfer. 2017;104:288-300.
[14] Wang CA, Sadat H, Tan JY. Meshless Method for Solving Transient Radiative and Conductive Heat Transfer in Two-Dimensional Complex Geometries. Numerical Heat Transfer Part B-Fundamentals. 2014;65:518-36.
[15] Li E, He ZC, Tang Q, Zhang GY. Large time steps in the explicit formulation of transient heat transfer. International Journal of Heat and Mass Transfer. 2017;108:2040-52.
[16] Bathe K-J. Finite element procedures: Klaus-Jurgen Bathe; 2006.
[17] Rieder C, Kroger T, Schumann C, Hahn HK. GPU-based real-time approximation of the ablation zone for radiofrequency ablation. IEEE transactions on visualization and computer graphics. 2011;17:1812-21.
[18] Kröger T, Altrogge I, Preusser T, Pereira PL, Schmidt D, Weihusen A, et al. Numerical Simulation of Radio Frequency Ablation with State Dependent Material Parameters in Three Space Dimensions. Berlin, Heidelberg: Springer Berlin Heidelberg; 2006. p. 380-8.
[19] Mariappan P, Weir P, Flanagan R, Voglreiter P, Alhonnoro T, Pollari M, et al. GPU-based RFA simulation for minimally invasive cancer treatment of liver tumours. Int J Comput Assist Radiol Surg. 2017;12:59-68.
[20] Wu S, Liu G, Zhang H, Xu X, Li Z. A node-based smoothed point interpolation method (NS-PIM) for three-dimensional heat transfer problems. International Journal of Thermal Sciences. 2009;48:1367-76.

[21] Liu P, Cui X, Deng J, Li S, Li Z, Chen L. Investigation of thermal responses during metallic additive manufacturing using a “Tri-Prism” finite element method. International Journal of Thermal Sciences. 2019;136:217-29.
[22] Yang K, Jiang GH, Li HY, Zhang ZB, Gao XW. Element differential method for solving transient heat conduction problems. International Journal of Heat and Mass Transfer. 2018;127:1189-97.
[23] Li Z, Cui X, Cai Y. Analysis of heat transfer problems using a novel low-order FEM based on gradient weighted operation. International Journal of Thermal Sciences. 2018;132:52-64.
[24] Yang T, Cui X. A random field model based on nodal integration domain for stochastic analysis of heat transfer problems. International Journal of Thermal Sciences. 2017;122:231-47.
[25] Ding CS, Cui XY, Deokar RR, Li GY, Cai Y, Tamma KK. An isogeometric independent coefficients (IGA-IC) reduced order method for accurate and efficient transient nonlinear heat conduction analysis. Numerical Heat Transfer Part a-Applications. 2018;73:667-84.
[26] Feng SZ, Cui XY, Li AM. Fast and efficient analysis of transient nonlinear heat conduction problems using combined approximations (CA) method. International Journal of Heat and Mass Transfer. 2016;97:638-44.
[27] Niroomandi S, Alfaro I, Cueto E, Chinesta F. Model order reduction for hyperelastic materials. International Journal for Numerical Methods in Engineering. 2010;81:1180-206.
[28] Zhang J, Zhong Y, Gu C. Deformable Models for Surgical Simulation: A Survey. IEEE Rev Biomed Eng. 2018;11:143-64.
[29] Wang H, Chong H, Li GY. Extension of reanalysis method, analysis of heat conduction by using CA and IC methods. Engineering Analysis with Boundary Elements. 2017;85:87-98.
[30] Zhang JN, Zhong YM, Gu CF. Energy balance method for modelling of soft tissue deformation. Computer-Aided Design. 2017;93:15-25.
[31] Rong X, Niu R, Liu G. Stability Analysis of Smoothed Finite Element Methods with Explicit Method for Transient Heat Transfer Problems. International Journal of Computational Methods. 2018:1845005.
[32] D’Alessandro V, Binci L, Montelpare S, Ricci R. On the development of OpenFOAM solvers based on explicit and implicit high-order Runge–Kutta schemes for incompressible flows with heat transfer. Computer Physics Communications. 2018;222:14-30.
[33] Ma J, Sun YS, Li BW. Spectral collocation method for transient thermal analysis of coupled conductive, convective and radiative heat transfer in the moving plate with temperature dependent properties and heat generation. International Journal of Heat and Mass Transfer. 2017;114:469-82.
[34] Cotin S, Delingette H, Ayache N. A hybrid elastic model for real-time cutting, deformations, and force feedback for surgery training and simulation. Visual Comput. 2000;16:437-52.
[35] Hauth M, Etzmuss O, Strasser W. Analysis of numerical methods for the simulation of deformable models. Visual Comput. 2003;19:581-600.
[36] Courant R, Friedrichs K, Lewy H. On the partial difference equations of mathematical physics. IBM journal of Research and Development. 1967;11:215-34.
[37] Wu S, Peng X, Zhang W, Bordas S. The virtual node polygonal element method for nonlinear thermal analysis with application to hybrid laser welding. International Journal of Heat and Mass Transfer. 2013;67:1247-54.
[38] Cui X, Li Z, Feng H, Feng S. Steady and transient heat transfer analysis using a stable node-based smoothed finite element method. International Journal of Thermal Sciences. 2016;110:12-25.
[39] Feng S, Cui X, Li A, Xie G. A face-based smoothed point interpolation method (FS-PIM) for analysis of nonlinear heat conduction in multi-material bodies. International Journal of Thermal Sciences. 2016;100:430-7.